\documentclass[twocolumn]{aastex62}

\usepackage{xcolor}
\usepackage{soul}
\graphicspath{{./}{figures/}}

\received{January 19, 2018}
\revised{January 19, 2018}
\accepted{\today}
\submitjournal{ApJ}

\shorttitle{Class I methanol maser database}
\shortauthors{Ladeyschikov et al.}

\begin{document}

\title{Online Database of Class I Methanol Masers}

\correspondingauthor{Dmitry Ladeyschikov}
\email{dmitry.ladeyschikov@urfu.ru}

\author{Dmitry A. Ladeyschikov}
\affil{Ural Federal University, 51 Lenin Str., 620051 Ekaterinburg, Russia}

\author{Olga S. Bayandina}
\affiliation{Joint Institute for VLBI ERIC, Oude Hoogeveensedijk 4, 7991 PD Dwingeloo, The Netherlands}

\author{Andrey M. Sobolev}
\affiliation{Ural Federal University, 51 Lenin Str., 620051 Ekaterinburg, Russia}



\begin{abstract}

In this paper, we present a database of class I methanol masers. The compiled information from the available literature provides \added{an} open and fast access to the data on class I methanol maser emission, including search, analysis and visualization of the extensive maser data set. 
\replaced{The  compiled information contains single-dish and interferometric data from published observations of class I methanol masers with detailed information about individual maser components from single-dish data and maser spots from interferometric data}{There is information on individual maser components detected with single-dish observations and maser spots obtained from interferometric data.}
At the moment the database contains information from $\sim$100 papers, \added{i.e.} $\sim$7500 observations and $\sim$650 sites of class I methanol masers. Analysis of the data \replaced{compiled}{collected} in the database shows that the distribution of class I methanol maser sources is similar to that of class II methanol masers. They are mostly \replaced{located}{found} in the Molecular Ring, where \replaced{most}{majority} of the OB stars are located. 
The difference between class I and II distributions is the presence of many class I methanol masers in the Nuclear Disk region (Central Molecular Zone). Access to the class I methanol maser database is available online at \href{http://maserdb.net}{http://maserdb.net}.

\end{abstract}

\keywords{stars: formation --- masers -- radio lines: ISM --- catalogs}


\section{Introduction}
Methanol maser emission is one of the features of star-forming regions. In early studies \replaced{by}{of} \citet{Batrla1987} and \citet{Menten1991}, two classes of methanol masers were empirically distinguished. \added{While} class II masers (for example, at 6.7, 12, 37.7, 107 GHz) are pumped by \deleted{the} infrared radiation of dust heated by young stars \citep{Sobolev2005,Cragg2005}\added{,} class I masers (for example, at 36, 44, 95 GHz) occur as a result of collisional-radiative pumping \citep{Sobolev2007} and usually indicate the presence of gas compressed by a shock wave. Such gas is often found near young stars, in which there is an outflow interacting with the surrounding material \citep{Voronkov2006}. These outflows \deleted{are} produc\replaced{ing}{e} shocks creating conditions for the pumping of class I methanol masers \citep{Bachiller1996}. Usually class I masers are found at some separation from young stellar objects  \citep[e.g.][]{Kurtz2004,Voronkov2014}. However, methanol masers of I class can occur anywhere in the interstellar medium where moderate-velocity shock waves are formed: in collisions of molecular gas clouds \citep{Salii2002}, at the boundaries of supernova remnants \citep{Pihlstrom2014}, in places where HII regions are interacting with the surrounding molecular gas  \citep{Voronkov2010b} and regions with complex hydrodynamic interactions \citep{Voronkov2010a}.

The first detection of the methanol \added{maser} emission was \deleted{one} in \added{the} $J_2 \rightarrow J_1$ ($J$=2..7) series of  methanol lines \replaced{at}{near} 25~GHz \deleted{by} \citep{Barrett71} and at 36~GHz by \citet{Turner72}. \replaced{Then the number of known methanol transitions with maser effect was increased}{Following these early observations the number of known methanol transitions showing maser emission has increased to five to the end of 1990, including transitions at 25, 36, 44, 84 and 95~GHz. Amount of known methanol maser sources also increased, as seen in the lower panel of the Figure~\ref{fig:eps} } 

At present \deleted{in the astronomy}, there is a trend to use large-scale sky surveys to solve various astrophysical problems instead of studying individual objects. This is a natural consequence of the fact that recently astronomers have gained access to a large number of sky surveys. The field of study of the class I methanol masers is not an exception. The increased sensitivity of single-dish and interferometric telescopes leads to a \replaced{large}{massive} amount of data that needs to be ordered.

In recent years \deleted{the latest} surveys of the northern and southern skies have \replaced{appeared}{been done} in the class I methanol maser lines. For example, the ATCA telescope as part of the MALT-45 project in 2015 produced a $``$blind$"$  overview of 5 square degrees in the southern sky at 7 mm \citep{Jordan2015}, including \deleted{in} the class I methanol line at 44 GHz. \replaced{77}{Seventy-seven} maser sources were registered, including 58 new detections.

\begin{deluxetable*}{llccl} 
\tablecaption{Known detected class I methanol maser transitions, included to the database of class I methanol masers.\label{tbl:transitions}}
\tablecolumns{5}
\tablehead{
\colhead{Transition} & \colhead{Short freq.} & \colhead{Rest frequency\tablenotemark{A}}  & \colhead{Detected\tablenotemark{B}} & \colhead{References\tablenotemark{C}\tablenotemark{D}}  \\
\colhead{} & \colhead{} & \colhead{MHz}  & \colhead{objects} & \colhead{} }
\colnumbers
\startdata
9$_{-1}$ $\rightarrow$ 8$_{-2}$~E & 9.93 GHz & 9936.202 [1] & 5 &   SLY93$^*$, VOR06, \textbf{VOR10B}, VOR11 \\
10$_{1}$ $\rightarrow$ 9$_{2}$~A$^{-}$ & 23.4 GHz  & 23444.759 [2] & G357.97-0.16, & VOR11 \\
 &   &  & G343.12-0.06 &  \\
$J_2$ $\rightarrow$ $J_1$~E Series & $\sim$25-30 GHz & --  & 23 & BAR71$^*$, MAT80, MEN86, JOH92, BEU05, \\
& &  & & VOR05, VOR06, VOR11, \textbf{TOW17} \\
4$_{-1}$ $\rightarrow$ 3$_{0}$~E & 36 GHz  & 36169.265 [1] &  210 &  MOR85$^*$, \textbf{LIE96}, \textbf{VOR14}, \textbf{COT16}\\ 
& & & & \textbf{BRE19}... 28 total \\
7$_{0}$ $\rightarrow$ 6$_{1}$~A$^{+}$ & 44 GHz  & 44069.410 [1] &  404 & MOR85$^*$, \textbf{SLY94}, \textbf{VOR14},\textbf{JOR15}, \\ 
& & & & \textbf{GOM16}, \textbf{JOR17}, \textbf{KIM19}... 45 total \\
5$_{-1}$ $\rightarrow$ 4$_{0}$~E & 84 GHz  & 84521.169 [1] &  129  &     BAT88$^*$,  \textbf{KAL01}, WIE04, \\ 
& & & &    KAL06, VOR06, \textbf{BRE19}\\
8$_{0}$ $\rightarrow$ 7$_{1}$~A$^{+}$ & 95 GHz  & 95169.463 [1]  & 534 & NAK86$^*$, \textbf{VAL95}, \textbf{VAL00}, \textbf{CHE11},\\ 
& &  & & \textbf{CHE12}, \textbf{GAN13}, \textbf{YAN17}... 25 total \\
11$_{-1}$ $\rightarrow$ 10$_{-2}$~E & 104.3 GHz  & 104300.414 [1] & W33-Met,  &  VOR04$^*$, VOR06 \\
 &   &  & IRAS16547-4247 & \\
6$_{-1}$ $\rightarrow$ 5$_{0}$~E & 132.8 GHz  & 132890.692 [1] & 16 &  SLY97$^*$, VAL98, SLY99, WIE04, \\
& & & &  CHO12, KAN13, LYO14   \\
4$_{2}$ $\rightarrow$ 3$_{1}$~E & 218.4 GHz  & 218440.063 [2] & NGC  6334I(N),  &      HUN14$^*$, CHE19 \\
 &  &  &  G352.630-1.067 &      \\
8$_{-1}$ $\rightarrow$ 7$_{0}$~E & 229.7 GHz  & 229758.756 [2] & 6 &      VAL98$^*$, SLY02, HUN14, CHE19 \\
5$_{2}$ $\rightarrow$ 4$_{1}$~E & 266.8 GHz  &  266838.148 [2] & G352.630-1.067  &      CHE19$^*$ \\
9$_{-1}$ $\rightarrow$ 8$_{0}$~E & 278.3 GHz  & 278304.512 [2]  & G34.43+00.24 MM3  &      YAN14$^*$
\enddata
\tablenotetext{A}{~Rest frequencies are taken from: [1] -- \citet{Muller2004}, [2] -- CDMS database \citep{Muller2005}}
\tablenotetext{B}{~In the case of \added{rare transitions with only} 1-2 detected objects, we \replaced{give}{list} the names of the objects. Maser objects \replaced{is}{are} maser groups with positive maser detection and maximum separation \added{of} 60 arcsec between detections. Details of maser grouping are presented in the Section~\ref{sect_group}. We \replaced{do}{did} not \replaced{count}{include} $\sim$2200 detections of 36 GHz methanol masers in the direction of CMZ from \citet{COT16} \replaced{to}{in} the total number of objects with 36 GHz masers. }
\tablenotetext{C}{~\added{Paper's} short names are coded as following: BAR71 -- \citet{BAR71}, BAT88 -- \citet{BAT88}, BEU05 -- \citet{BEU05}, BRE19 -- \citet{BRE19}, CHE11 -- \citet{CHE11}, CHE12 -- \citet{CHE12}, CHE19 -- \citet{CHE19}, CHO12 -- \citet{CHO12}, COT16 -- \citet{COT16}, GAN13 -- \citet{GAN13}, GOM16 -- \citet{GOM16}, HUN14 -- \citet{HUN14}, JOH92 -- \citet{JOH92}, JOR15 -- \citet{JOR15}, JOR17 -- \citet{JOR17}, KAL01 -- \citet{KAL01}, KAL06 -- \citet{KAL06}, KAN13 -- \citet{KAN13}, LIE96 -- \citet{LIE96}, LYO14 -- \citet{LYO14}, MAT80 -- \citet{MAT80}, MEN86 -- \citet{MEN86}, MOR85 -- \citet{MOR85}, NAK86 -- \citet{NAK86}, SLY02 -- \citet{SLY02}, SLY93 -- \citet{SLY93}, SLY94 -- \citet{SLY94}, SLY97 -- \citet{SLY97}, SLY99 -- \citet{SLY99}, TOW17 -- \citet{TOW17}, VAL00 -- \citet{VAL00}, VAL95 -- \citet{VAL95}, VAL98 -- \citet{VAL98}, VOR04 -- \citet{VOR04}, VOR05 -- \citet{VOR05}, VOR06 -- \citet{VOR06}, VOR10B -- \citet{VOR10B}, VOR11 -- \citet{VOR11}, VOR14 -- \citet{VOR14}, WIE04 -- \citet{WIE04}, KIM19 -- \citet{KIM19}, YAN14 -- \citet{YAN14}, YAN17 -- \citet{YAN17}.}
\tablenotetext{D}{~Papers with asterisks (*) are \added{the} first maser emission detections of \deleted{particular} methanol transition. Papers \deleted{marked} with \replaced{bold face}{boldface} are \replaced{large}{extensive} surveys of \deleted{particular} methanol maser\replaced{transition}{s}.
}
\end{deluxetable*}

In the northern sky, a 13.7-meter telescope at the Observatory of Purple Mountain (China) \replaced{conducted a search}{searched} for class I methanol masers in the direction of $\sim$1000 molecular clumps at 95~GHz \citep{Yang2017}. As a result, 205 sources with maser emission were detected.

In the last years, the amount of the interferometric studies of methanol maser sites increased significantly \added{(see the lower panel of Figure~\ref{fig:eps})}, providing a detailed view of methanol maser sources with information on the individual maser spots. In the last \replaced{10}{ten} years the number of class I methanol interferomet\added{r}ic  studies increase\added{d} from 11 to 39 papers. All these works are evidence that observations of methanol masers are currently relevant and are regularly carried out to study the star formation fields. Creating a database will solve the problem of access to the observational data of the interferometric studies.

Until now, work on creating a catalog of class I methanol masers was carried out by Astro Space Center group\footnote{Astro Space Center of Lebedev Physical Institute of RAS} \citep{Valtts2007,Valtts2010,Bayandina2012}. This group presents a catalog of 206 objects, obtained from the literature up to 2011 inclusive. This catalog can be used as a starting point for the further systematization of accumulated data to achieve the completeness of data collection on class I methanol masers. Another starting point is the Avedisova catalog of star-forming regions \citep{Avedisova2002}. This work is a unique study on the systematization of information about star formation regions for papers up to 2001, including class I methanol masers.

In the present paper, we introduce an improved data access interface and the \deleted{most} complete set of data on class I methanol masers, which will allow \replaced{performing}{doing} \replaced{their}{the} statistical analysis, \replaced{carrying out studies}{study of the} individual objects and \replaced{planning}{plan for the} future observations. \replaced{Class II masers are also in the field of our maser database, but the process of data input for class II methanol masers is not complete to date. Nevertheless, we use the currently on-going results of class II maser database to compare Galactic distribution of class I and class II methanol masers,}{We are in the process of including the complete sample of known class II methanol masers into the database and while this effort is not yet completed, we use this information to compare Galactic distribution of class I and class II methanol masers}, as described in the Section~\ref{classII}. Rather than focusing on class II masers\added{,} we primarily analyze the class I sources that are currently insufficiently studied. Detailed analysis of class II masers is available in the literature \citep[e.g.][]{Pestalozzi07,Pandian07,Breen11,Green17}.

\deleted{In} Our database \replaced{, we include}{contains} not only most widespread class I transitions (25, 36, 44, 84, 95 GHz), but also relatively rare transitions \deleted{which were} identified as class I methanol maser emission in the discovery papers. These include $6_{-1}\rightarrow 5_0$~E transition (132.8~GHz) with 16 known objects, $9_{-1}\rightarrow 8_{-2}$~E transition (9.93~GHz) with 8 known objects and $8_{-1}\rightarrow 7_0$~E transition (229.7 GHz) with 6 detected objects. The recent detections of rare transitions include $10_{1}\rightarrow 9_2$~A$^-$ (23.4~GHz) in G357.97-0.16 \added{and 343.12-0.06}  \citep{VOR11}, $11_{-1}\rightarrow 10_{-2}$~E (104.3~GHz) in W33-Met \citep{VOR04} and IRAS 16547-4247 \citep{VOR06}, $4_{2}\rightarrow 3_1$~E (218.4 GHz) in NGC 6334I(N) \citep{HUN14} and G352.630-1.067 \citep{CHE19}, $5_{2}\rightarrow 4_1$~E (266.8~GHz)  in G352.630-1.067 \citep{CHE19} and $9_{-1}\rightarrow 8_0$~E (278.3~GHz) in G34.43+00.24 MM3 \citep{YAN14}. The full list of detected class I maser transitions included to the database is given in the Table~\ref{tbl:transitions}.

Class I methanol masers were detected not only in our \deleted{own} Galaxy but in several external galaxies. More than 25 years ago, \citet{Sobolev1993} suggested that search for extragalactic methanol masers should be most promising in the starburst galaxies in the 36.2 GHz class I methanol transition. After that\added{,} there \replaced{was}{were} \replaced{a number of}{several} unsuccessful searches for methanol masers in galaxies \citep{Ellingsen1994,Phillips1998,Darling2003}, but these searches \replaced{was}{were} focused on the class II transition at 6.7 GHz, probably inspired by the detection of 6.7 GHz methanol maser in LMC \citep{Sinclair1992}. Only in 2014 methanol maser emission was detected at 36.2 GHz (class I) toward the nearby starburst \replaced{G}{g}alaxy NGC 253 \citep{ELL14}. After this detection\added{,} NGC 253 became the target of extensive interferometric studies \citep{ELL17,GOR17,CHE18,MCC18B}. Extragalactic Class I methanol maser at 84 GHz was first detected in Seyfert 2 galaxy NGC 1068 \citep{WAN14}. Later methanol masers at 36 and 37 GHz (class I and II) were detected in the ultra-luminous infrared galaxy Arp 220 \citep{CHE15}. \citet{CHE16} \added{report tentative detections} of class I methanol masers in 11 galaxies associated with water and OH megamasers. Recently class I methanol maser at 36~GHz was also detected in NGC 4945 \citep{MCC17,MCC18}. To date, class I methanol masers were detected in 14 galaxies \added{(11 of these were tentative detections)}.

\section{Data processing}

\subsection{Data entry}
Data entry to the database was carried out in semi-automatic mode with the support of a special information system, \replaced{written}{developed} specifically for this project. In the case \replaced{where}{when} the authors \deleted{of some particular papers} publish the source data in the electronic format, \replaced{these}{this} source data were used. If it was possible to obtain data from the authors \deleted{themselves} with their permission, \replaced{also}{then} data was entered \added{directly} into the database. In other cases, tabular data was converted from the PDF format to CSV using OCR technology (Optical Character Recognition). If the access to the article is restricted, then PDF files from a repository of electronic preprints arXiv (\hyperlink{https://arxiv.org/}{https://arxiv.org}) are used. Together with tabular data for each article, textual descriptions of sources and images \deleted{, created by the authors of the articles,} were entered into the database. \replaced{All these}{The} data was entered into the information system, which identifie\replaced{s}{ed}, cross-referenc\replaced{e}{ed} this information and prepare\replaced{s}{ed} it for the \replaced{entry}{entrance} into the database. Maser data\added{base} \replaced{is stored in}{works on} the PostgreSQL database \replaced{on}{in} the Ural Federal University virtual machine service. For sky queries, we \replaced{use}{used} PgSphere extension \citep{Chilingarian2004} for PostgreSQL, that allow doing the cone search, x-match and \deleted{do} other tasks with sky coordinates.

The entered data \replaced{are}{is} divided into three main classes: targeted \added{single-dish} surveys, blind surveys, and \added{targeted}  interferometric \replaced{follow-ups}{observations}. 
The database stores in table \textit{met1\_data} all useful information about observations that is available in papers, including maser emission properties (radial velocity, intensity, line width), size and position angle of the beam, the date of observations and the name of the facility \deleted{on which the observations were made}. Additional information for interferometric data about detected maser spots is stored in a separate table \textit{maser\_spots} of the database.

To search for articles on class I methanol masers, the new NASA ADS system\footnote{\hyperlink{http://ui.adsabs.harvard.edu}{https://ui.adsabs.harvard.edu}} was queried using\replaced{key words}{keywords} (See Appendix \ref{Appendix1}). Then this list was filtered by hands: we keep only articles with unique observations of class I methanol maser sources. This procedure is periodically repeated to identify new papers containing observations of class I methanol masers. Further, this list was divided into the three main types of papers -- targeted single-dish surveys, blind surveys and \added{targeted} interferometric observations. Each type of paper has unique features of the data-entry process. For \added{targeted} single-dish and blind surveys, only two tables \deleted{are} needed: source table with coordinates and table with the line parameters. For \added{targeted} interferometric data, one more table is essential: maser spot data.

\subsection{Search for the class I methanol masers data}
In total\added{,} there are about $\sim$100 papers \deleted{that are included} in our database. We divide all papers into three different groups -- targeted \added{single-dish} surveys, blind surveys, and \added{targeted} interferometric \deleted{follow-up} observations. 

Targeted \added{single-dish} surveys in our classification \replaced{are distinguished by absence of}{have no} information about individual spots of maser sources, but \replaced{presence of}{have the} criteria of the source selection. Blind surveys \added{include both single-dish and interferometric data} and have no criteria of the source selection \replaced{and}{but} only have sky region defined. \replaced{Follow-up}{Targeted} interferometric observations have information about maser spots and have some criteria of the source selection.

For further classification, for each paper we create the following labels: \texttt{data\_type} is comma-separated list of detected molecules in the paper, \texttt{frequencies} is comma-separated shortlist of frequencies which were observed in the paper, \texttt{obs\_type} is type of paper from our classification (targeted single-dish \added{survey}, blind survey or \added{targeted} interferometric observations), \texttt{target\_type} is \replaced{classification of way of targeting}{category of target} in the particular paper. For polarization observations, we additionally add label "polarization" to the \texttt{obs\_type} column. The full list of included papers \added{is} presented in the Table~\ref{tbl:papers} of Appendix~\ref{Appendix1} and the summary on the target objects list \added{is} presented in the Table~\ref{tbl:stat}.

\subsubsection{Targeted single-dish surveys}
Targeted surveys are the most popular way to search for class I methanol masers. \added{This category includes single-dish surveys with some source selection criteria.} The tendency is that the more recent papers target sources from the large sky catalogs (GLIMPSE, Bolocam). In the older surveys, the sources \added{were} primarily selected using the known list of high-mass star formation regions and class II methanol masers. \added{This can be illustrated in the lower panel of Figure~\ref{fig:eps}. Starting from 2010 the number of known maser sites with detected class~I maser emission grown very fast.} \added{To date} the most extensive single-dish surveys \replaced{for the present time is done}{are} \replaced{by}{of} \citet{YAN17} with 205 detections at 95 GHz and \added{of} \citet{CHE11} with 105 detections at 95 GHz.

\subsubsection{Blind surveys}
Blind surveys in class I methanol line is another way to search for the new class I methanol masers. \added{This category includes both single-dish and interferometric surveys without source selection criteria.} The main goal of these observations was to find all detectable sources in the selected sky area. We found only five papers that use this observing technique for class I methanol masers - \citet{COT16,JOR15,LYO14,YUS13,VOR11}. 

In the work of \citet{COT16} over 900 points were observed to cover a region  $66'\times13'$ along the inner Galactic plane, providing the catalog of 2240 methanol maser spots in the CMZ at 36~GHz. The MALT-45 survey \citep{Jordan2015} observed 5 square degrees ($330^{\circ}<l<335^{\circ}$, b = $\pm$ 0.5$^{\circ}$) for spectral lines in the 7 mm band, including class I methanol maser transition at 44 GHz. \citet{Lyo2014} mapped the $7'\times10'$  area of NGC 1333 in 22 GHz H$_2$O and 44 GHz CH$_3$OH maser transition. \added{An} extensive search for the class I methanol masers with frequencies \added{of} about 25 GHz was \added{conducted as} part of the HOPS project \citep{Walsh2011} \replaced{which}{and} lead to discovery of the 23.4~GHz class~I maser methanol emission in G357.97-0.16  \citep{VOR11}. \added{23.4~GHz maser was also detected in the same paper toward the source 343.12-0.06 using the ATCA follow-up observations}.

\subsubsection{Targeted interferometric observations}
\added{This category includes interferometric observations with some source selection criteria.} Our database stores more than 40 papers with interferometric \deleted{follow-up} observations, including \replaced{7}{seven} papers with interferometic studies of the external galaxies. 

The most extensive survey\added{s} in our sample are \citet{Jordan2017} with 238 maser spots detected in 77 \added{maser} sites at 44~GHz and \citet{Voronkov2014} with more than 700 maser spots detected in 71 \deleted{maser} sites \replaced{in}{at} 36 and 44 GHz \deleted{maser transitions}. 

\subsection{Catalogue description}
Each entry of the main catalogue \texttt{met1\_data} contains information about a single spectral component of methanol maser at specific position in sky and radial velocity. The following columns are available for each entry: observation identifier (\texttt{obs\_id}); source name (\texttt{source\_name}); source identifier using short galactic coordinates (\texttt{grp}) in format G0.000$ \pm $0.000 (N), where N is a number of available observations of this source in different papers; right ascension and declination in J2000 coordinates (\texttt{ra} and \texttt{dec}) in degrees; galactic longitude and latitude (\texttt{l} and \texttt{b}) in degrees; detection flag (\texttt{detected}); peak velocity (\texttt{vpeak}); FWHM of line (\texttt{fwhm}) in km/s; peak intensity (\texttt{peak}) in Jy or K units; integrated intensity (\texttt{int}) in Jy km/s or K km/s units; type of units for intensity (\texttt{units}: 0 for Jy and 1 for K); distance to the source in kpc (\texttt{dist}); columns for beam information (\texttt{beam}, \texttt{beam\_maj}, \texttt{beam\_min} in arcsec, \texttt{beam\_pa} in degrees); root mean square noise of observation  (\texttt{rms}) and units of rms (\texttt{units\_rms}: 0 for Jy and 1 for K); date of observation in arbitrary format (\texttt{date\_txt}); reference to the paper (\texttt{ref}); short name of the facility (\texttt{telescope\_ref}); description of the source from paper (\texttt{descr}) and link to the associated author's image from the paper (\texttt{image}). 

Additionally, there are a few flags for each observation: \texttt{gc\_flag} reveal Central Molecular Zone sources ($359.3 <$ \texttt{l} $< 1.7$ and $\vert$\texttt{b}$\vert$ $<0.2$); \texttt{mjy\_flag} and \texttt{rms\_mjy\_flag} shows mJy units instead of Jy for intensity and rms; \texttt{is\_interf} shows that the observation has interferometric \deleted{follow-up} data for individual maser spots; \texttt{thermal} denote lines which were described as thermal in the original papers. The columns \texttt{iras}, \texttt{twomass}, \texttt{wise}, \texttt{akari}, \texttt{akari\_fis}, \texttt{bgps}, \texttt{continuum} and \texttt{ego} show associated sources in popular astronomical catalogs, which are useful in class I methanol maser studies. If \replaced{the}{a} source is observed simultaneously in several methanol lines, then each observation will be included to the database with the corresponding \texttt{line} column.

For \added{targeted} interferometric \replaced{follow-ups}{observations} the data \replaced{about}{on} individual masers spots is stored in the separate table \texttt{maser\_spots}. The format of the table columns is following: \texttt{obs\_id} is reference to the observation identifier in \added{the} main table \texttt{met1\_data}; \texttt{spot\_number} is sequential number of maser spot \added{detected} in the \deleted{single interferometeric observation} field; \texttt{source\_name}\ is \replaced{name}{an identifier} of  each maser spot composed of \replaced{using combination}{a} main source name and letters A..Z,A1..Z1,..; \texttt{spot\_location} is \added{an} optional field of maser spot location (usually comes from the source paper); right ascension and declination in J2000 coordinates (\texttt{ra} and \texttt{dec}) in degrees; galactic longitude and latitude (\texttt{l} and \texttt{b}) in degrees; width of line at half maximum (\texttt{linewidth}); peak velocity in km/s (\texttt{vpeak}); peak intensity (\texttt{peak}) in Jy (mJy) or K (mK), where units specified in the main table in the corresponding entry with specific \texttt{obs\_id}; \texttt{ref} is the reference paper; \texttt{offset\_ra} and \texttt{offset\_dec} \replaced{is}{are} relative \added{offsets} (in milliarcseconds) \deleted{coordinates} of the maser spot \added{position}  \replaced{relative to main}{from} the target coordinates of the source, specified in the main table with corresponding \texttt{obs\_id}. 

\subsection{Data completeness}
The presented database is complete for known to date class I methanol masers sources. We check\added{ed} database completeness using the list of papers produced with NASA ADS system (see Appendix~\ref{Appendix1} for SQL-query that we used for searching the papers) and list of the methanol class I maser line catalogs available at Strasburg archive\footnote{\hyperlink{http://vizier.u-strasbg.fr/viz-bin/VizieR}{http://vizier.u-strasbg.fr/viz-bin/VizieR}}. We found no papers published since 1990 \replaced{that}{and} contain\added{ing} original observations of class I methanol masers that were not included in our database. \replaced{For papers older than 1990 there are sometimes not enough information about maser observation (missing peak velocities or intensities , etc).}{Papers published before the 90s often miss important observational information, such as peak velocities or intensities}. 
In that case, we do not include information from these \deleted{older} papers to our database if the \deleted{observed} source \added{is} well-known and more recent observations of class I methanol line \replaced{is}{are} available \deleted{in these sources}. \replaced{However, we force to include old paper where the first detection of some particular methanol maser transition was done}{However, we include older papers where the first detection of methanol maser transitions were made}.

\subsection{Class II methanol masers} \label{classII}
We use incomplete database of class II methanol masers that is currently under construction\footnote{Access to the class II maser database is the same as for class I database: \href{http://maserdb.net}{http://maserdb.net}} to compare the distributions of Class I and Class II methanol maser sources in the Galaxy. To date, the class II maser database already contains $\sim8500$ observations in $\sim3400$ source groups (with $\sim4600$ detections in $\sim1200$ source groups) and provides extensive information on the overall distribution of class II masers. The database contains most recent \deleted{large} extensive surveys of class II masers, including the full 6.7 GHz methanol multibeam maser catalogue \citep[e.g.][]{Breen15}, the 6.7 GHz methanol maser survey with the Shanghai Tianma Radio Telescope \citep{Yang17,Yang19}, the Torun catalogue of 6.7 GHz methanol masers \citep{Szymczak12}, the general catalogue of 6.7 GHz methanol masers \citep{Pestalozzi05} and others. \added{The} total amount of papers which are currently covered is 35. 

\subsection{Source grouping}\label{sect_group}

\begin{figure}
	\plotone{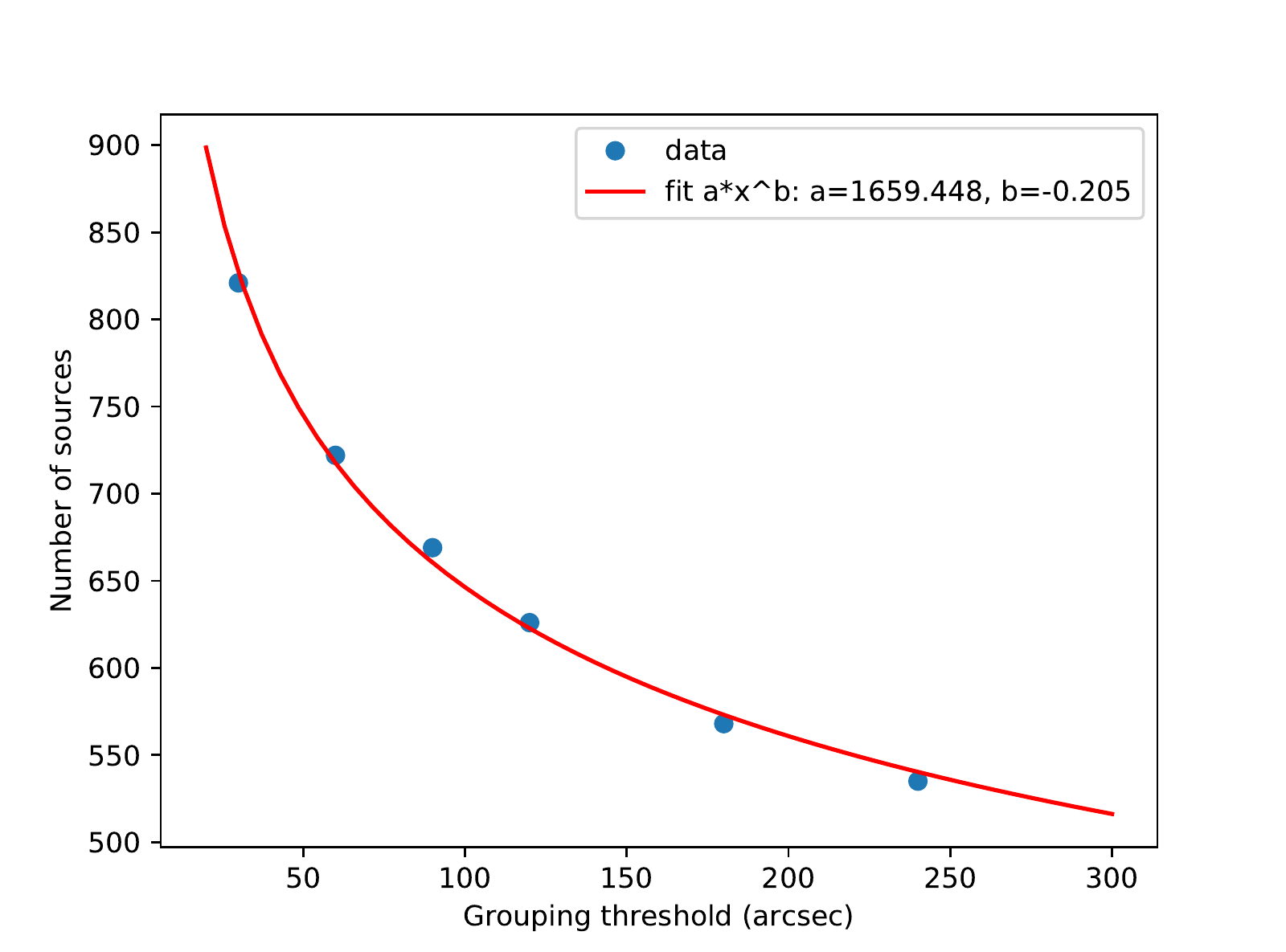}
	\plotone{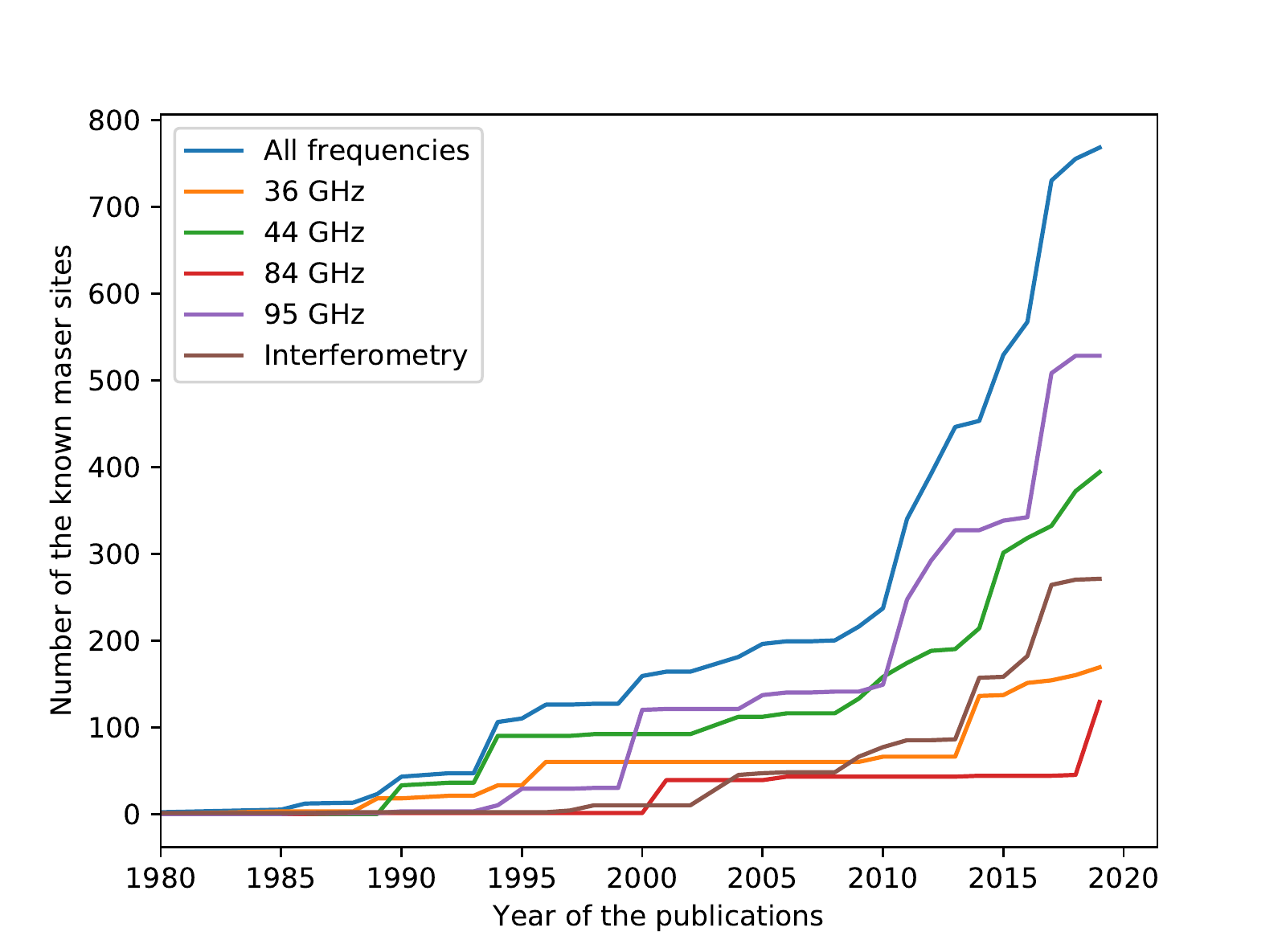}
    \caption{Upper panel: \replaced{Dependence of $\varepsilon$ threshold of \texttt{DBSCAN} algorithm to the resulting number of detected maser sites.}{Relation of the resulting number of detected maser sites from $\varepsilon$ threshold of \texttt{DBSCAN} algorithm.} \added{Lower panel: dependence of the number of known maser sites from the publication date.} }
    \label{fig:eps}
\end{figure}

The main catalog table of maser observation contains individual maser detection and non-detection entries from each paper. Most of the sources are observed many times in the different papers, often in the different methanol transitions. If we need to know the total number of individual sources and do the statistical analysis of maser detection, we need to combine the available observations into source groups. 

We perform the source grouping using a density-based clustering algorithm \texttt{DBSCAN}, firstly proposed by \citet{Ester1996}. The main advantage of this clustering algorithm for our database is that \replaced{one to specify the number of clusters in the data a priori}{it does not require a priori specialization of the number of clusters in the data}. We use \texttt{SCIKIT-LEARN} python \citep{Pedregosa2012} realization of \texttt{DBSCAN}. The algorithm requires two input parameters for clustering: $\varepsilon$ for minimal separation between cluster points and $\mathrm{N}_{min}$ for \added{the} minimal number of cluster points (we assume $\mathrm{N}_{min}=1$). The different values of $\varepsilon$ leads to a different number of source groups. \replaced{From this figure, we can see that the number of source groups is 735-566 when using values of $\varepsilon=60-120$ arcsec.}{Figure~\ref{fig:eps} displays the dependence of the resulting number of source groups from $\varepsilon$}. In the final version of the database, we assume $\varepsilon=60$ arcsec with the resulting number of sources of 735. 

\replaced{When the source groups are considered,  algorithm does not use the distances to the observed sources.}{Note that the algorithm does not consider distances to sources when calculating source groups.} 
 The source groups are created using only the angular separation between sources in the sky. \added{Therefore,} some detections may be combined in one object even if they are not physically related. \replaced{This is up to}{The} researcher \added{has} to distinguish between different physical objects. We include the information about the distances to \deleted{of the observed} a source to the database if it is available in the original papers.

Each group receive it's name based on the median average galactic coordinates in the form of $``$G0.000$\pm$0.000$~(N)"$, where $N$ is the number of maser observations included in the source group. From the grouping process we create the table \texttt{met1\_objects}, where each entry is the individual source object. The description of the columns in this table is following: \texttt{group\_name} is shorten galactic coordinates of the group with number of observations in brackets; \texttt{source\_name} is the \deleted{first} name of the source from the main table; \texttt{source\_names} is semicolon-separated list of source names of this object from literature; \texttt{ra} and \texttt{dec} \replaced{is}{are} median mean of right ascension and declination in J2000 coordinates in degrees; \texttt{l} and \texttt{b} \replaced{is}{are} median mean of galactic longitude and latitude in degrees; \texttt{vpeak} is semicolon-separated list of peak velocities of the all detected spectral components; \texttt{vmean} is the median average of the peak velocities; \texttt{det} is detection flag of I class methanol line (Y/N); \texttt{line} is semicolon-separated list of detected class I lines in the format \added{e.g.} \texttt{8(0)->7(1)\_A+}; \texttt{gc\_flag} is CMZ indicator ($359.3 < l < 1.7$ and $\vert b\vert < 0.2$); \texttt{refs} is semicolon-separated list of papers where observations of this source is presented. Additionally, some catalog association columns are present: \texttt{iras}, \texttt{akari}, \texttt{akari}\_fis, \texttt{ego}, \texttt{continuum}, \texttt{bgps}. Using this table, we may investigate the list of all known observed and detected sources of class I methanol maser emission in different transitions.

\subsection{Association with the external catalogs}

Association with the external astronomical catalogs is a very efficient way to investigate the nature \added{of} objects. We select the following catalogs to study the class I methanol masers: \textit{IRAS} \citep{Helou1988}, \textit{Akari} \citep{Ishihara2010}, \textit{EGO} \citep{Cyganowski2008}, \textit{Bolocam GPS} \citep{Ginsburg2013}, and well-known \textit{continuum} catalogs \citep{White1992,Wright1994,Becker1991,Condon1998,Wright1990}. Since methanol masers of the class I usually appear at some distance from the associated objects, we do not include near- and mid-infrared point source catalogs (e.g. \textit{2MASS}, \textit{WISE}). Instead, we use only catalogs with a moderate number of sources where the association may be \replaced{done}{established} in automated cross-matching process. We use the web API of \texttt{CDS X-Match Service} \citep{Pineau2011} to process the matching.

The algorithm for finding associations is following. For each observation, we find the nearest sources in different catalogs within the search radius \added{of} 60 arcsec. If for a particular \replaced{observation}{source}, there are no sources within 60 arcsec, then we repeat the search\deleted{ing} using an increased radius of 72 arcsec. If there is no match within \deleted{catalog sources anyway are not found with} the increased radius, \added{a} particular source is considered to have no association in the catalog. If the association is found, then the catalog source name is stored in the corresponding field of the database.

In the case of source groups, that \added{were} determined by \texttt{DBSCAN} algorithm, all external sources, associated with some particular group are stored in the semicolon-separated list for each group.

Parameters of the associated catalog sources are stored in the local database, thus it is possible to quickly retrieve fluxes of the associated sources without accessing to the VizieR database and use them in further studies.

\section{Access to the class I methanol maser database}

The class I methanol maser database can be accessed via the web interface \added{with the link}  \href{http://maserdb.net}{http://maserdb.net}. In this section we examine the main functionality of the web interface.

\subsection{Main search page}

\begin{figure}
	\plotone{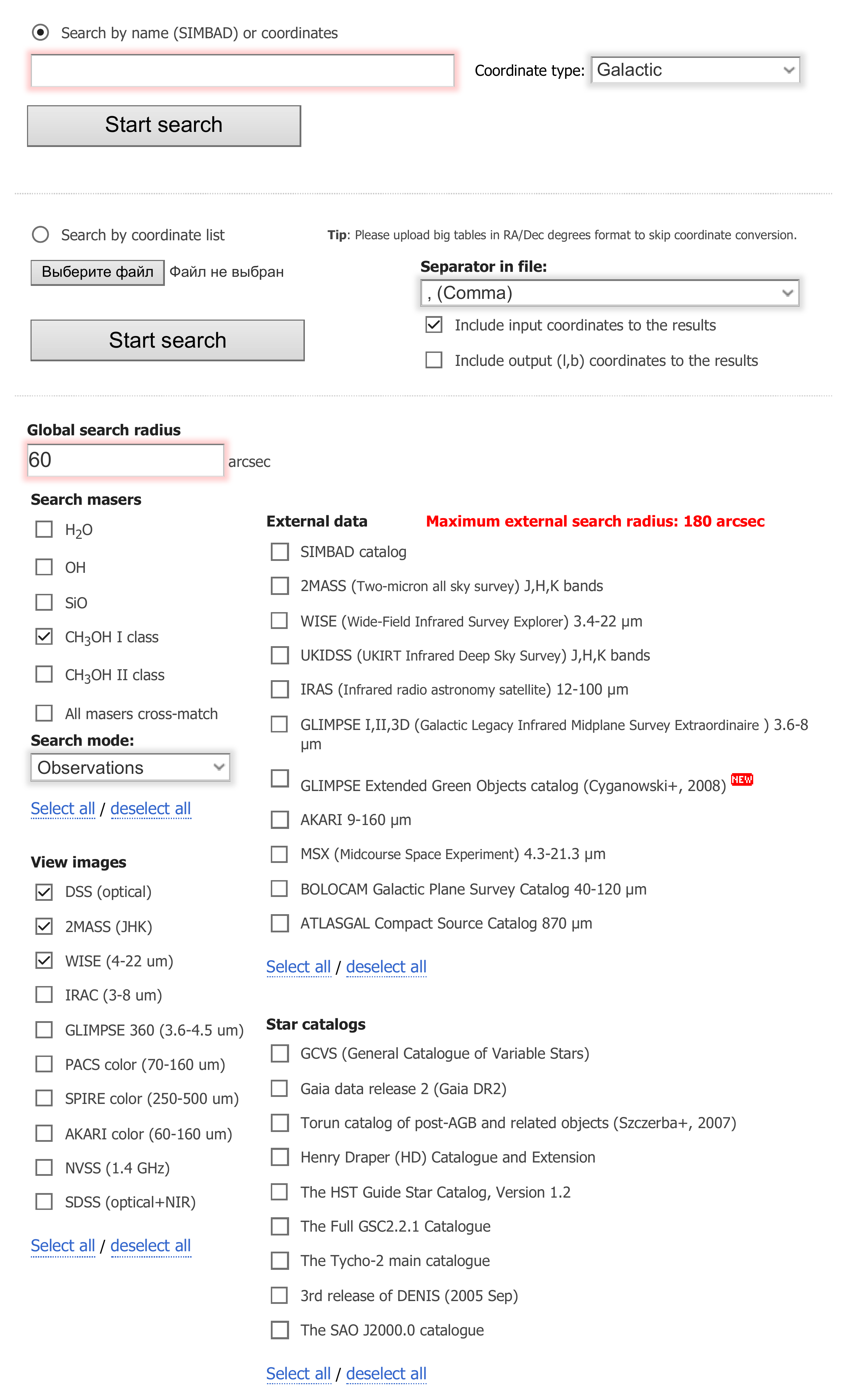}
    \caption{The main search page of the database web-interface.}
    \label{fig:search}
\end{figure}

The main \deleted{search} page of the web interface (see Figure~\ref{fig:search}) \replaced{provides the ability to search for maser observations and sources in different ways. There is a possibility to do the search in the following manners:}{provides several search options}:

\begin{itemize}
    \item \textbf{Cone search} by coordinates or source name. \replaced{If user inputs coordinates, then the coordinate system should be specified. The possible coordinate systems are Equatorial (J2000 or B1950) or Galactic. If the user inputs the source name, then the search of the corresponding Simbad source will be performed. If the source is not found, then the search in the local maser database will be done.}{The corresponding coordinate system, either Equatorial (J2000 or B1950) or Galactic, should be specified in the case of the coordinates search. External search of Simbad\footnote{\hyperlink{http://simbad.u-strasbg.fr/simbad/sim-fid}{http://simbad.u-strasbg.fr/simbad/sim-fid}} database will be performed in the case of the search by a source name.}

    \item \textbf{List search}. The search allows to input the list of sources (an example of an entry in a list: \textit{o Ceti; 02 19 20.7; -02 58 39}) to \added{perform} simultaneous search of masers and other sources \added{within} each list entry. 
    This may be useful for observers \replaced{with the list of sources where they need to know which sources have detected or non-detected in class I methanol maser emission.}{willing to test source lists on detection or non-detection of class I methanol maser emission in them.}
    
\end{itemize}\par

\noindent Optionally user may choose the search mode:

\begin{itemize}
    \item \textbf{Observations}: the search \deleted{results} returns all found observations \deleted{in the database} without grouping. Note that the list of detections could be very large for well-known sources.\par

    \item \textbf{Sources}: the search \deleted{results} returns only \deleted{the} source groups. In this mode, all detections are grouped using the algorithm described in the Section~\ref{sect_group}.
\end{itemize}\par

\added{The} web system allows access\replaced{ing}{to} \deleted{the} external data, including images and catalogs of well-known sky surveys. The visual inspection is done using Aladin Lite service\citep{Boch2014} and catalog search is done by CDS X-Match service \citep{Pineau2011}. The results of catalog and image search are displayed \added{at} the same search results window together with the $``$local$"$  maser database search results, allowing to examine the sources \replaced{in}{with} the different wavelengths. \replaced{As the database contains source descriptions and author images from the original papers, it is also displayed in the search results with the corresponding reference}{Source description and author images from the original papers are also displayed in the search results with the corresponding reference}.

\subsection{References page}

The page allows to browse and study the individual papers stored in the database. Click at each paper leads to the paper's page with title, abstract and full data listing. The quality of entered data may be checked in this page, comparing the online data with the author's original tables in PDFs -- the parameters of all observed sources are displayed in the data table. Viewing images and displaying the associated data from the well-known astronomical catalog is possible in this page. Each source in the data table is clickable: one can get more information about each observed source in some particular paper. For \deleted{interferometry} papers with \added{interferometric data}, the individual maser spots are displayed for each observed source. 

\replaced{There is also an ability}{It is also possible} to view the online paper statistics, including distribution of sources in the all-sky diagram, distribution of sources in IRAS color-color diagram, distribution of the detection velocities versus galactic longitude and distribution of sources in galactic latitude and longitude histograms. Examples of online figures are \replaced{the same as}{presented on} Figures~\ref{fig:posvel}.

\subsection{Objects page}

The objects page is the result of source grouping described in the Section~\ref{sect_group}. This page allows \replaced{to select}{selecting} of the full list of objects with the specific maser molecules detected (or only observed). For example, it \replaced{give an answer to}{answers} one of the following question: what sources have maser detection in both H$_2$O and class I methanol maser lines? Currently, not only the class I methanol maser emission is included in the database, but also class II methanol masers, H$_2$O, OH and SiO maser emission. However, the data entry process for these molecules is currently not finished.

\replaced{After}{When} the maser molecule is selected, the list of objects appears in the search result window. One can display the associated data (external catalogs) and view images of each object. The $``$Detection$"$  field in the list provide information about detected and not detected molecules in each object.

The object statistics page may also be displayed for selected objects similar to the one in the reference page.

\subsection{Download page}

This page gives the ability to get the local copy of the maser database in order to study the maser properties in the popular astronomical software, for example -- \texttt{TOPCAT} \citep{Taylor2005}. One can download data in two modes: observations or objects for each maser species. The object mode is the group of observations \replaced{into}{for} individual sources as described in the Section~\ref{sect_group}. The data is returned in CSV format. The observations mode is the original observations from the papers.

\begin{figure*}
\epsscale{1.2}
\plotone{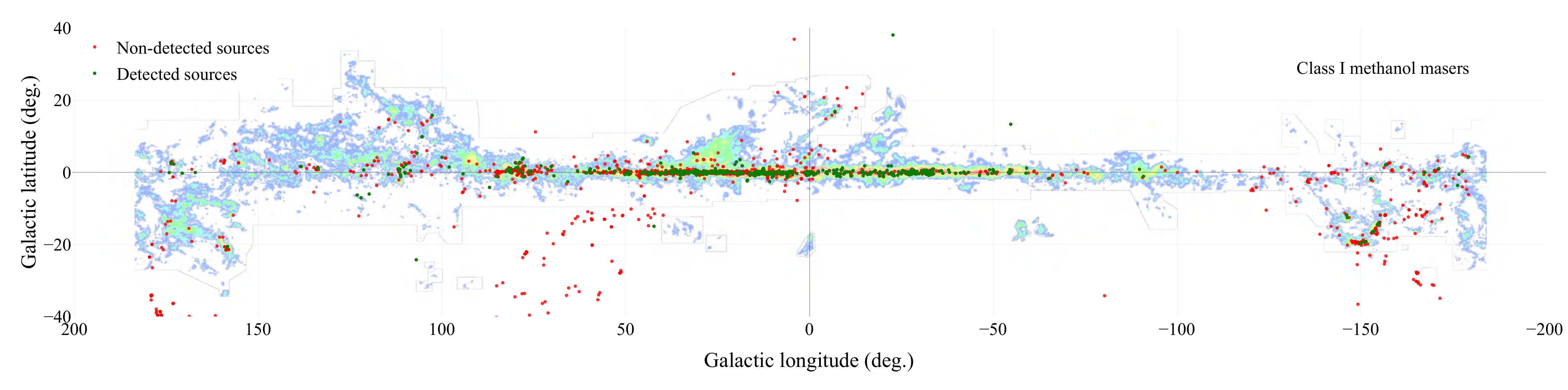}
\plotone{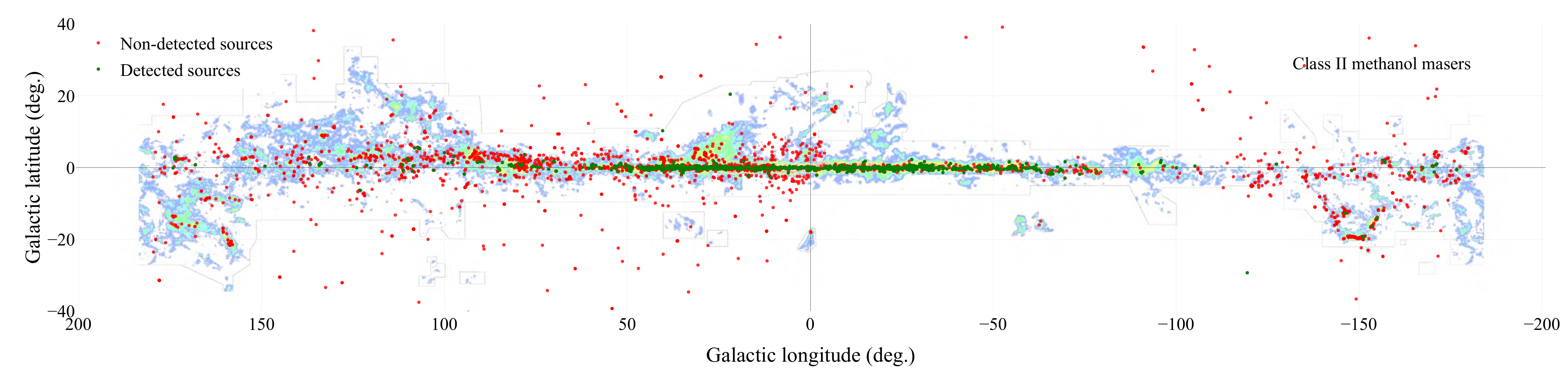}
\plotone{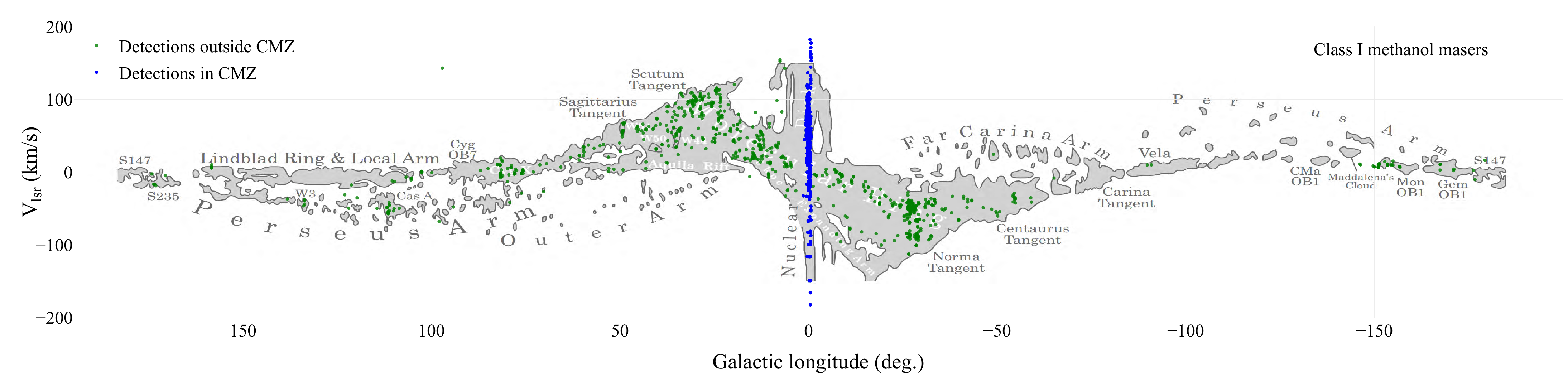}
\plotone{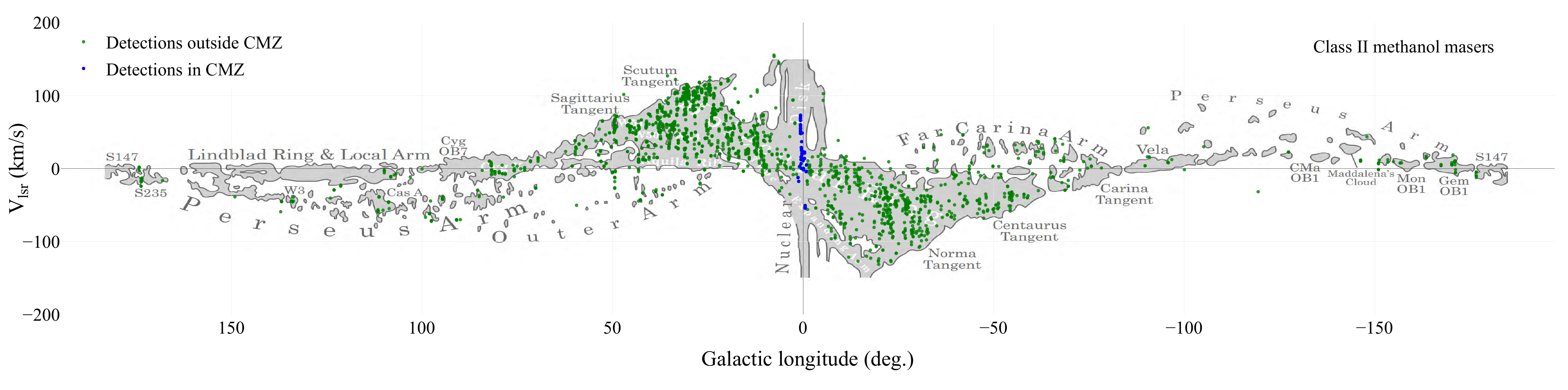}
    \caption{Distribution of class I/II methanol masers in the Position-Position and Position-Velocity space. Background CO image\added{s} are from \citet{Dame2001}. Green and red dots in \added{the} top two panels are detections and non-detections\added{, respectively}. Blue and green dots in two bottom panels are detections inside and outside of the Central Molecular Zone ($359.3 <$ \texttt{l} $< 1.7$ and $\vert$\texttt{b}$\vert$ $<0.2$)\added{, respectively}.}
    \label{fig:posvel}
\end{figure*}

\section{Discussion} \label{discussion}

We utilize the database to review the distribution of class I methanol masers in the position and velocity.  Position-Position diagram (see Figure~\ref{fig:posvel}) reveal\added{s} that methanol masers of class I \added{are} located primarily in the Galactic plane. 90$\%$  of detected class I methanol masers have $|b| < 1.5$ deg. The concentration of class I methanol masers increase to the direction of the Galactic center and decrease at the Galactic anti-center: 77$\%$  of detected class I methanol masers have $320 < l < 40$ deg. The same is reported in \citet{Yang2017} -- the authors notice that class I methanol masers  concentrate\deleted{d} in a central ring of $l = \pm40$ deg with a width of about 1.6 deg in galactic latitude. This concentration may be a result of the association of class I methanol masers with Galactic star-formation regions \deleted{, that is} located mainly in the Galactic plane.

Position-velocity diagram (see Figure~\ref{fig:posvel}) reveals that the \replaced{richest}{most populated} in class I methanol maser sites \replaced{is}{are} CMZ (Central Molecular Zone or Nuclear Disk) and Molecular Ring. It was recently discovered that the CMZ is rich in methanol emission \citep{Yusef2013} and contains \replaced{many}{a lot of} class I methanol masers \cite{COT16}. The Molecular Ring is the dominant structure in the distribution of the molecular gas, traced by CO emission \citep{Dobbs2012}, and it is also rich in class I methanol masers. Other class I methanol maser nurser\replaced{y}{ies} are primarily associated with star-formation regions, including S235, W3, Cygnus X, NGC 7538, Ori A $\&$  B, Mon OB1, Gem OB1, and others. A small number of detections (14 sources) are of extragalactic origin.

From Figure~\ref{fig:posvel} one can notice that the distribution of class I methanol masers in the Position-Velocity diagram similar to the distribution of massive star formation regions \citep{Bronfman2000}, where \added{the} most of the OB associations are located. This is consistent with the \replaced{view}{theory} that methanol masers are closely associated with star formation regions. However, we should note that there is a bias in the distribution of the detected class I masers, as the most popular 
\replaced{way of targeting for observation of class I masers is high-mass star formation regions}{category of target of class I maser observations is high-mass star formation regions}. Other classes of objects have \replaced{less}{a lower number of} observations, \replaced{as the}{consequently} detection rates toward these object classes are much lower \deleted{and they are not widespread in the observations}. In comparison, distribution of class II masers have much more observations toward objects other than high-mass star formation regions \deleted{(with non-detections)}, thus class II maser distribution is less biased \replaced{in comparison with}{than} class I. \added{This could be seen} in the top two panels of the Figure ~\ref{fig:posvel}. 

\replaced{When}{We} \replaced{comparing}{compared} the distribution of class I and class II methanol masers in the Position-Velocity diagram (Figure~\ref{fig:posvel}). \added{O}ne can notice that the main difference between them is the \replaced{great}{high} number of class I methanol masers at the Central Molecular Zone (CMZ or Nuclear Disk in the figure of \citet{Dame2001}), and \added{the} much smaller number of class II masers there. At the same time, class II maser sources are distributed more widely along Molecular Ring in comparison to Class I masers.  The sources in the Nuclear Disk that seen in the Position-Velocity diagram come mainly from the works \replaced{by}{of} \citet{COT16}. The discovery of abundant $4_{-1}\rightarrow 3_0$~E methanol emission in CMZ was done in \citet{Yusef2013}. However, \added{low} velocity resolution \replaced{in the configuration of the telescope}{of the observation}  ($\sim$16.6~km~s$^{-1}$) was not \replaced{able}{sufficient} to distinguish \deleted{between} thermal and maser components. The authors explained the origin of \added{the methanol}  enhanced abundance \deleted{of methanol} in terms of the interaction of cosmic rays and molecular gas in the CMZ. The confirmation of \added{the} presence of class I methanol masers in CMZ was done in \cite{COT16}. The authors notice that the origin of such a large number of methanol masers is not clear, but it appears unlikely that the entire population of these masers trace the early phase of star formation in the Galactic center \citep{COT16}.

The pumping mechanism for class I methanol masers is different from class II \citep{Sobolev2007}. While class II sources \added{are} pumped mainly by radiation and \added{their} \replaced{connection between them and}{association with} star formation regions is tighter, class I methanol masers \added{are} pumped by collisions. They may \replaced{appear}{arise} not only in \deleted{the} star formation regions but in any place where gas \added{is} compressed by a shock wave. \added{The complex high-velocity gas interactions that lead to high-velocity features may explain a large number of \deleted{the} class I methanol masers in the CMZ}. \added{These gas interactions are} not associated with star formation. This may be similar to the detection of 36~GHz maser and thermal methanol emission at the periphery of the CMZ \citep{Menten2009,Salii2002}. In some rare cases, high-velocity features even may be found outside of the CMZ \citep{Voronkov2010a}, but the Galactic center region is the most active and dynamic \deleted{region} in our Galaxy, thus many class I methanol masers arise because of gas interactions.

\begin{deluxetable*}{llcccl} 
\tablecaption{Statistics of class I methanol maser detection \replaced{in}{with} the different \replaced{ways of targeting}{categories of targets}. Values in brackets give\deleted{s} the detection rates. References in italics are \added{targeted} interferometric \added{observations} papers. Explanation of the papers short references \added{is} given in the Table~\ref{tbl:papers} \label{tbl:stat}}
\tablecolumns{6}

\tablehead{
\colhead{Object type} & \colhead{Detected} & \colhead{Non-det.}  & \colhead{Single-dish} & \colhead{Interf.} & \colhead{References' short names} \\
\colhead{} & \colhead{Objects} & \colhead{Objects}  & \colhead{Papers} & \colhead{Papers} & \colhead{} 
}
\colnumbers
\startdata
Bolocam (BGPS) sources & \replaced{470}{468} (34\%) & 899 & 2 & 0 & YAN17,CHE12 \\
Young Stellar Objects & 172 (31\%) & 375 & 5 & 5 & \textit{GOM16},\textit{HUN14},KAN13,LIM12,BAE11,\textit{CHA11},\added{KIM15} \\
 &  &  &  &  &  CHE11,\textit{CYG09},\textit{VOR06},KAL92 \\
Molecular Outflows & \replaced{163}{170} (\replaced{31}{32}\%) & \replaced{364}{368} & 3 & 2 & \added{KIM15}, GAN13,CHE11,\textit{KAL10},\textit{ARA08} \\
Known Class I methanol masers & \replaced{261}{260} (51\%) & 251 & 11 & 3 & BRE19,\textit{MCC18C},\textit{JOR17},KAN16,KAN15,\textit{VOR14}, \\
searched for new transitions &  &  &  &  & ELL05,WIE04,KAL01,VAL00,ROL99,LIE96, \\
 &  &  &  &  & SLY94,BER92 \\
GLIMPSE sources & 158 (41\%) & 231 & 2 & 1 & CHE12,CHE11,\textit{CYG09} \\
Known Class II methanol masers & 152 (39\%) & 233 & 6 & 1 & \textit{MCC18C},KAN15,ELL05,WIE04,VAL00,ROL99,SLY94 \\
High-mass star formation regions & 133 (58\%) & \replaced{97}{101} & 11 & 18 & \textit{CHE19},\textit{ELL18},\textit{TOW17},\textit{ROD17},\textit{MOM17},\textit{MAT14},LYO14, \\
 &  &  &  &  & CHO12,\textit{FIS11},\textit{VOR10B},FON10,\textit{GOM10}, \\
  &  &  &  &  & POL10,\textit{SLY09},\textit{ARA09},\textit{ARA08},\textit{VOR05},\textit{KUR04},SLY02, \\   
  &  &  &  &  & \textit{KOG98},\textit{LIE97},SLY97,VAL95,KAL94B,KAL94, \\
  &  &  &  &  & \textit{JOH92},PLA90,HAS90,NAK86,\textit{MAT80} \\
Ultra-Compact HII region & 99 (49\%) & 105 & 3 & 2 & KIM19,\textit{GOM16},\textit{BEU05},KAL96,FOR90 \\
Extended Green Objects & 112 (55\%) & 92 & 1 & 2 & \textit{TOW17},CHE11,\textit{CYG09} \\
Low-mass star formation regions & 8 (11\%) & 67 & 1 & 3 & \textit{KAL13},\textit{KAL10},\textit{KAL10B},KAL06 \\
HII regions & 44 (59\%) & 30 & 2 & 1 & \textit{ROD17},HAS89,MOR85 \\
Post AGB/Planetary nebulas & 0 (0\%) & 73 & 1 & 0 & GOM14 \\
Supernova remnants & 8 (31\%) & 18 & 3 & 2 & LI17,\textit{MCE16B},\textit{PIH14},LIT11,ZUB08 \\
Other & 19  & 1 & 4 & 1 & NAK15,LIM12,\textit{VOR10},ZUB07,VAL98 \\
Extragalactic object & 14 & 4 & 3 & 7 & \textit{MCC18B},\textit{CHE18},\textit{MCC18},\textit{GOR17},\textit{ELL17},MCC17, \\
  &  &  &  &  &CHE16,\textit{CHE15},WAN14,\textit{ELL14} \\
CO emission clumps & 4 (40\%) & 6 & 1 & 0 & LAD16 \\
Infrared Dark Clouds & 2 & 0 & 1 & 1 & \textit{YAN14},DEG12B \\
Central Molecular Zone & $\sim$2000 &  & 2 & 3 & COT16,YUS13,\textit{SJO11},\textit{PIH11},\textit{MEH97} 
\enddata
\end{deluxetable*}

\subsection{Statistics of maser detection in the different categories of targets}

We analyze the statistics of detections and non-detections \replaced{of}{associated with} the different \replaced{ways of targeting}{categories of targets} using the groups of sources, described in the Section~\ref{sect_group}. The results are presented in the Table~\ref{tbl:stat}. The \replaced{classes of the way of targeting}{categories of targets} are marked for each paper based on object selection criterion in the original paper. In some papers, objects have been selected using several \replaced{ways of targeting}{categories of targets}, listed in the Table~\ref{tbl:stat}. For example, \citet{Chen2011} targets a sample of YSOs from the GLIMPSE survey based on the color characteristics of EGOs with ongoing outflows. In this case, the paper was included in the following \replaced{ways of targeting}{categories of targets}: \texttt{EGO, GLIMPSE, YSO, Outflows}. The full list of papers with \replaced{corresponding way of targeting}{similar categories of targets} \added{is} presented in the Table~\ref{tbl:papers}.

We notice that different \replaced{ways of targeting}{categories of targets} may have \added{an} intersection. Most intersections occur\deleted{s} in the targeting to the Molecular Outflows (166 objects), Young Stellar Objects (179 objects), and Extended Green Objects (112 objects). If we combine these samples into one, it will be only 223 sources in total, but simple summing up gives 457 sources. The intersection occurs because these  samples consist of the sources manifesting different phenomena found in the same kind of objects -- young stellar objects. EGOs are objects displaying enhanced emission in 4.5 $\mu$m band of Spitzer IRAC  \citep{Cyganowski2008}. Enhanced emission is caused by H$_2$ line and CO (v=1-0) bandhead emission, which falls into 4.5 $\mu$m band and may be excited by the outflow activity of young stellar objects. It is also reported that there may be false EGO detections \citep[e.g.][]{DeBuizer2010}. However, YSOs are not always associated with EGOs detected by Spitzer. \citet{LIM12} found 282 new objects of diffuse/collimated H$_2$ emission in UWISH2 survey (UKIRT Widefield Infrared Survey for H$_2$) at 2.122 $\mu$m. This UWISH2 2.122 $\mu$m data is 300--2000 times deeper in terms of the H$_2$ emission sensitivity than the corresponding Spitzer IRAC data. From that sample $\sim$5\% of sources have associated class I methanol masers at 44~GHz confirming their YSO origin. All these objects are not detected as known EGOs or star formation regions \citep{LIM12}. Similar situation takes place with molecular outflows: most EGOs trace population of actively accreting YSO outflow candidates \citep{Cyganowski2008}, but molecular outflows are not always associated with enhanced 4.5 $\mu$m band emission (EGOs) in Spitzer data and may be traced \replaced{by}{with} other \replaced{methods}{markers}, for example CO emission \citep{Wu2004}, H$_2$ emission at 2.122 $\mu$m \citep{Varricatt2010,Froebrich2011}. In our study, we did not classify objects and only \replaced{marked}{specified} classes of the objects provided in the publications.

The most popular \replaced{way of targeting}{category of targets} to search for the class I methanol masers is \added{the} survey of the high-mass star formation regions (HMSFRs) -- there are in total 29 papers (11 papers based on single-dish and 18 papers based on interferometric data) with 133 objects detected \replaced{in}{by}  class I methanol maser emission. 
The \replaced{largest}{most significant} number of class I methanol maser sources \replaced{were}{was} detected toward the sources from the Bolocam Galactic Plane Survey -- 470 maser sites \citep{Yang2017,Chen2012}. The detection rate for Bolocam sources is about 34\%. In the second place by the number of new detections \replaced{are}{is} a search for the new maser transitions in the sources \replaced{which}{that} already display\deleted{ed} class I methanol maser emission in some other transition (261 detected objects).
Another fruitful \replaced{way of targeting}{category of targets} brought detection of  class I methanol masers in 158 objects by looking at the sources from the Spitzer GLIMPSE survey \citep{CHE12,CHE11,CYG09}. Search for class I methanol masers in the samples of known class II methanol masers brought detection of 152 class I methanol masers \citep[e.g.][]{ELL05, KAN15}. 

The detection rates for the different \replaced{ways of targeting}{categories of targets} range from 16\% to 58\%. The highest detection rate is toward the high-mass star formation regions (58\%), class I maser sources searched for new transitions (51\%), Extended Green Objects (55\%), HII regions (59\%) and Ultra-compact HII regions (49\%). The lowest detection rate is for the low-mass star formation regions (11\%). The detection rate for the blind survey of CMZ \citep{COT16,YUS13} cannot be estimated. 

\deleted{In} The most extensive search for extragalactic class I masers of \citet{CHE16} aimed \deleted{at} the extended X-ray galaxies \deleted{which} show\added{ing} megamaser emission of water or OH. blue{It was} expect\replaced{ing}{ed} that the detection rate of class I methanol masers could be higher \replaced{in}{toward} the sources with megamasers. They found 11 galaxies with class I masers from their sample of 16 objects, thus formal detection rate is 68\% which is very high. \added{However, these detections are reported to be tentative}.

\section{Conclusions}

In this paper, we present the database of class I methanol masers \replaced{which has}{with} the following characteristics:

\begin{itemize}
    \item Data is \replaced{compiled}{collected} from \deleted{the} literature and is constantly updated.
    
    \item The total number of observations of class I methanol masers included in the database is $\sim$7500 entries (\added{with} $\sim$9500 \replaced{when counting}{of} individual maser components)
    
    \item Interferometric data on maser spots is given in the separate table \textit{maser\_spots} and contains $\sim$15000 observations of maser spots in $\sim$460 sources.
    
    \item The database contains tabular data \replaced{for all}{from all papers published on the topic to date}, \added{with} descriptions of the observed sources from 24 papers (starting from 1998) and original author's images from 46 papers (starting from 1992).
\end{itemize}

\noindent From \replaced{the}{an} analysis of methanol maser database, we found that:

\begin{itemize}
    \item There are in total $\sim$650 maser sites with positive detection of class I methanol masers, not including the CMZ (Central Molecular Zone) sources. Number of maser sites depends on the way of grouping and range from 735 to 566 when using different separation threshold for grouping, ranging from 60 to 180 arcsec.\par

    \item In the Position-Velocity diagram most of class I methanol masers are associated with the sources \replaced{situated in}{hosted by} Molecular Ring, Central Molecular Zone and Galactic Star Formation Regions (S235, W3, Cygnus X, NGC 7538, Ori A $\&$ B, Mon OB1, Gem OB1 and others). Class I methanol masers are detected in 14 external galaxies.\par

    \item The distribution of class I methanol masers in Position-Velocity diagram is similar to that of class II masers in almost all structures of the Galaxy displaying molecular emission. \replaced{The only exclusion is}{With exception of} CMZ, where numerous class I methanol masers were detected by \citet{COT16} with no evidence of association with \added{the} formation of massive stars. The origin of \added{a} such large number of class I methanol masers in CMZ is currently not clear.
    
    \item From the statistical analysis, we found that the most popular \replaced{way of targeting}{category of targets} for the search of class I methanol masers \replaced{uses}{is} high-mass star formation regions  \deleted{as a targets} (29 papers with 133 objects detected). The \replaced{largest}{most significant} number of class I methanol maser sources was detected in \added{the} Bolocam survey (472 objects). The highest detection rates are toward the samples of high-mass star formation regions (58\%), known class I methanol masers in search for new maser transitions (59\%), Extended Green Objects (55\%), HII regions (56\%) and Ultra-compact HII regions (57\%). Detections of class I methanol maser sources are highly dominated by the high-mass star formation regions and only a few sources \deleted{are} found in the low-mass star-forming sites.
\end{itemize}\par


\section*{Acknowledgements}

The work of DAL, including input of the data on class I methanol masers in the database, analysis of class I methanol maser's distribution in the Galaxy and statistical analysis of the survey detection rates were funded by Russian Foundation for Basic Research through research project 18-32-00605. Studies of relative distributions of class I and class II methanol masers supported by Russian Science Foundation grant 18-12-00193. The work was supported in part by the Ministry of Education and Science of Russia (the basic part of the Stateassignment, RK no. AAAA-A17-117030310283-7).

\appendix
\section{List of papers} \label{Appendix1}

The Table~\ref{tbl:papers} provide\added{s} \replaced{the}{a} full list of papers that \added{were} included in the class I methanol maser database \replaced{at the moment of}{to the date of} August 2019. The database is updated regularly, thus new papers may be added. The online version of this list \replaced{may be found}{is posted} on the website of the database\footnote{\hyperlink{http://maserdb.net/list.pl}{http://maserdb.net/list.pl}}.

The following query for ADS system was used in order to find papers, containing class I methanol masers data: \texttt{(abs:"class I" OR abs:"36 GHz" OR abs:"44 GHz" OR abs:"95 GHz" OR abs:"25 GHz" OR abs:"84 GHz" OR abs:"9.9 GHz" OR abs:"104 GHz") AND (title: "methanol" OR title: "CH3OH")}

\startlongtable

\begin{deluxetable*}{llcccccc}
\tabletypesize{\footnotesize}
\tablecaption{Papers with original observations that was included inthe class I maser database. The number of detection and non-detection are counted from the source groups and may differ from the original paper's source count. Asterisks (*) mark the papers with the combination of different categories of targets.\label{tbl:papers}}
\tablehead{
\colhead{id} & \colhead{Ref} & \colhead{Short title}  & \colhead{Target type} & \colhead{Det.} & \colhead{Non-det.} & Frequencies & \colhead{Mode} 
}
\colnumbers
\startdata
1 & KIM19 & \citet{KIM19} & UCHII & 55 & 67 & 44 & Single-dish \\
2 & CHE19 & \citet{CHE19} & HMSFR & 1 & 0 & 218,230,267 & Interf. \\
3 & BRE19 & \citet{BRE19} & Class\_I(36,44) & 92 & 2 & 36,85 & Single-dish \\
4 & MCC18 & \citet{MCC18} & Extragalactic & 1 & 0 & 36 & Interf. \\
5 & MCC18B & \citet{MCC18B} & Extragalactic & 1 & 0 & 36,85 & Interf. \\
6 & MCC18C & \citet{MCC18C} & Class\_I(95),Class\_II(6) & 30 & 0 & 95 & Interf. \\
7 & ELL18 & \citet{ELL18} & HMSFR & 11 & 0 & 36 & Interf. \\
8 & CHE18 & \citet{CHE18} & Extragalactic & 1 & 0 & 36 & Interf. \\
9 & ELL17 & \citet{ELL17} & Extragalactic & 1 & 0 & 36 & Interf. \\
10 & GOR17 & \citet{GOR17} & Extragalactic & 1 & 0 & 36 & Interf. \\
11 & YAN17 & \citet{YAN17} & Bolocam & 462 & 760 & 95 & Single-dish \\
12 & TOW17 & \citet{TOW17} & HMSFR,EGO & 12 & 8 & 25 & Interf. \\
13 & ROD17B & \citet{ROD17B} & HMSFR,HII & 24 & 30 & 44 & Interf. \\
14 & MOM17 & \citet{MOM17} & HMSFR & 1 & 0 & 44 & Interf. \\
15 & MCC17 & \citet{MCC17} & Extragalactic & 1 & 0 & 36 & Single-dish \\
16 & LI17 & \citet{LI17} & SNR & 1 & 0 & 95 & Single-dish \\
17 & JOR17 & \citet{JOR17} & Class\_I(44) & 66 & 9 & 44 & Interf. \\
18 & COT16 & \citet{COT16} & CMZ & 34 & 0 & 36 & Blind \\
19 & MCE16B & \citet{MCE16B} & SNR & 3 & 1 & 36,44 & Interf. \\
20 & LAD16 & \citet{LAD16} & CO\_clumps,IRAS & 4 & 6 & 36 & Single-dish \\
21 & KAN16 & \citet{KAN16} & Class\_I(44) & 36 & 0 & 44,95 & Single-dish,\\
 &  & &  &  &  &  & Polarization \\
22 & GOM16 & \citet{GOM16} & UCHII,YSO & 29 & 40 & 44 & Interf. \\
23 & CHE16 & \citet{CHE16} & Extragalactic & 10 & 4 & 36 & Single-dish \\
24 & NAK15 & \citet{NAK15} & Other & 1 & 1 & 23,36,44,95,109 & Single-dish \\
25 & KAN15 & \citet{KAN15} & Class\_II(6) & 25 & 55 & 44,95 & Single-dish \\
26 & JOR15 & \citet{JOR15} & Blind & 73 & 0 & 44 & Blind \\
27 & CHE15 & \citet{CHE15} & Extragalactic & 1 & 0 & 36 & Interf. \\
28 & PIH14 & \citet{PIH14} & SNR & 2 & 18 & 36,44 & Interf. \\
29 & MAT14 & \citet{MAT14} & HMSFR & 1 & 0 & 44 & Interf. \\
30 & LYO14 & \citet{LYO14} & HMSFR & 2 & 0 & 44,95,133 & Blind \\
31 & HUN14 & \citet{HUN14} & YSO,Discovery & 1 & 0 & 218,230 & Interf. \\
32 & ELL14 & \citet{ELL14} & Extragalactic & 1 & 0 & 36 & Interf. \\
33 & GOM14 & \citet{GOM14} & pAGB\_PN & 0 & 73 & 25 & Single-dish \\
34 & YAN14 & \citet{YAN14} & IRDC,Discovery & 1 & 0 & 278 & Interf. \\
35 & WAN14 & \citet{WAN14} & Extragalactic & 1 & 0 & 85 & Single-dish \\
36 & VOR14 & \citet{VOR14} & Class\_I(36,44,95) & 71 & 0 & 36,44 & Interf. \\
37 & KAL13 & \citet{KAL13} & LMSFR & 4 & 0 & 44 & Interf. \\
38 & YUS13 & \citet{YUS13} & CMZ,Discovery & 41 & 0 & 36 & Blind \\
39 & KAN13 & \citet{KAN13} & YSO & 4 & 53 & 44,95,133 & Single-dish \\
40 & GAN13 & \citet{GAN13} & Outflows & 55 & 222 & 95 & Single-dish \\
41 & CHE12 & \citet{CHE12} & GLIMPSE,Bolocam & 63 & 150 & 95 & Single-dish \\
42 & LIM12 & \citet{LIM12} & YSO,Other & 13 & 0 & 44 & Single-dish \\
43 & DEG12B & \citet{DEG12B} & Other & 1 & 0 & 44 & Single-dish \\
44 & CHO12 & \citet{CHO12} & HMSFR & 1 & 1 & 23,44,87,95,133 & Single-dish \\
45 & BAE11 & \citet{BAE11} & YSO & 10 & 176 & 44 & Single-dish \\
46 & CHA11 & \citet{CHA11} & YSO & 6 & 22 & 44 & Interf. \\
47 & CHE11 & \citet{CHE11} & EGO,GLIMPSE,& 103 & 87 & 95 & Single-dish \\
 &  & & YSO,Outflows &  &  &  & \\
48 & FIS11 & \citet{FIS11} & HMSFR & 3 & 0 & 36 & Interf. \\
49 & LIT11 & \citet{LIT11} & SNR & 1 & 0 & 44 & Single-dish \\
50 & LIT11B & \citet{LIT11B} & IRAS,UCHII,YSO,& 9 & 54 & 44 & Single-dish \\
 &  &  & Outflows,HII* &  &  &  &  \\
51 & PIH11 & \citet{PIH11} & CMZ & 1 & 0 & 44 & Interf. \\
52 & SJO11 & \citet{SJO11} & CMZ & 1 & 0 & 36 & Interf. \\
53 & VOR11 & \citet{VOR11} & Discovery & 1 & 0 & 10,23,25 & Single-dish \\
54 & FON10 & \citet{FON10} & HMSFR,IRAS & 27 & 0 & 44,95 & Single-dish \\
55 & VOR10B & \citet{VOR10B} & HMSFR & 4 & 45 & 10 & Interf. \\
56 & VOR10 & \citet{VOR10} & Other & 1 & 0 & 36,44 & Interf. \\
57 & POL10 & \citet{POL10} & HMSFR & 3 & 0 & 36,44 & Single-dish \\
58 & KAL10 & \citet{KAL10B} & LMSFR,Outflows & 7 & 67 & 36,44 & Interf. \\
59 & KAL10B & \citet{KAL10} & LMSFR & 1 & 0 & 44 & Interf. \\
60 & GOM10 & \citet{GOM10} & HMSFR & 3 & 0 & 44 & Interf. \\
61 & CYG09 & \citet{CYG09} & EGO,GLIMPSE,YSO & 18 & 1 & 44 & Interf. \\
62 & SLY09 & \citet{SLY09} & HMSFR & 1 & 2 & 44,95 & Interf. \\
63 & ARA09 & \citet{ARA09} & HMSFR & 1 & 0 & 44 & Interf. \\
64 & ARA08 & \citet{ARA08} & HMSFR,Outflows & 1 & 0 & 44 & Interf. \\
65 & ZUB08 & \citet{ZUB08} & SNR & 1 & 0 & 95 & Single-dish \\
66 & ZUB07 & \citet{ZUB07} & Other & 3 & 0 & 95 & Single-dish \\
67 & VOR06 & \citet{VOR06} & YSO & 1 & 1 & 10,25,26,85,95,104 & Interf. \\
68 & KAL06 & \citet{KAL06} & LMSFR & 4 & 0 & 44,85,95 & Single-dish \\
69 & ELL05 & \citet{ELL05} & Class\_II(6) & 25 & 31 & 95 & Single-dish \\
70 & VOR05 & \citet{VOR05} & HMSFR & 1 & 0 & 25 & Interf. \\
71 & BEU05 & \citet{BEU05} & UCHII & 1 & 1 & 25 & Interf. \\
72 & WIE04 & \citet{WIE04} & Class\_I(84,95,133),& 10 & 0 & 85,95,133 & Single-dish,\\
 &  & & Class\_II(107,157) &  &  &  &  Polarization \\
73 & KUR04 & \citet{KUR04} & HMSFR & 36 & 10 & 44 & Interf. \\
74 & SLY02 & \citet{SLY02} & HMSFR & 6 & 6 & 109,165,230 & Single-dish \\
75 & KAL01 & \citet{KAL01} & Class\_I & 39 & 3 & 85,95 & Single-dish \\
76 & VAL00 & \citet{VAL00} & Class\_I(36,44),Class\_II & 100 & 68 & 95 & Single-dish \\
77 & ROL99 & \citet{ROL99} & Class\_II(6) & 0 & 87 & 10 & Single-dish \\
78 & KOG98 & \citet{KOG98} & HMSFR & 8 & 0 & 44 & Interf. \\
79 & VAL98 & \citet{VAL98} & Other & 1 & 0 & 7,12,44,95,107,& Single-dish \\
 &  &  &  &  &  & 109,133,157,230 & \\
80 & SLY97 & \citet{SLY97} & HMSFR & 6 & 0 & 133 & Single-dish \\
81 & MEH97 & \citet{MEH97} & CMZ & 1 & 0 & 44 & Interf. \\
82 & LIE97 & \citet{LIE97} & HMSFR & 2 & 0 &  & Interf. \\
83 & LIE96 & \citet{LIE96} & Class\_I(25,44,84,95) & 38 & 7 & 36 & Single-dish \\
84 & KAL96 & \citet{KAL96} & UCHII & 17 & 0 & 36 & Single-dish \\
85 & VAL95 & \citet{VAL95} & HMSFR & 27 & 0 & 95 & Single-dish \\
86 & KAL94B & \citet{KAL94B} & HMSFR & 8 & 0 & 95 & Single-dish \\
87 & SLY94 & \citet{SLY94} & Class\_II(6) & 58 & 0 & 44 & Single-dish \\
88 & KAL94 & \citet{KAL94} & HMSFR & 16 & 0 & 36 & Single-dish \\
89 & SLY93 & \citet{SLY93} & Discovery & 3 & 3 & 10 & Single-dish \\
90 & KAL93 & \citet{KAL93} & IRAS & 3 & 1 & 36,44 & Single-dish \\
91 & JOH92 & \citet{JOH92} & HMSFR & 1 & 0 & 25 & Interf. \\
92 & KAL92 & \citet{KAL92} & YSO,IRAS & 3 & 0 & 36,44 & Single-dish \\
93 & BER92 & \citet{BER92} & Class\_I & 8 & 0 & 36 & Single-dish \\
94 & FOR90 & \citet{FOR90} & UCHII & 10 & 0 & 44 & Single-dish \\
95 & PLA90 & \citet{PLA90} & HMSFR & 2 & 0 & 95 & Single-dish \\
96 & HAS90 & \citet{HAS90} & HMSFR & 19 & 1 & 44 & Single-dish \\
97 & BAC90 & \citet{BAC90} & H2O & 15 & 0 & 44 & Single-dish \\
98 & HAS89 & \citet{HAS89} & HII & 19 & 0 & 36 & Single-dish \\
99 & BAT88 & \citet{BAT88} & Discovery & 1 & 0 & 85 & Interf. \\
100 & MEN86 & \citet{MEN86} & Discovery & 4 & 0 & 25 & Single-dish \\
101 & MEN86B & \citet{MEN86B} & Discovery & 4 & 0 & 23 & Single-dish \\
102 & NAK86 & \citet{NAK86} & HMSFR & 1 & 0 & 95 & Single-dish \\
103 & MOR85 & \citet{MOR85} & Discovery,HII & 4 & 0 & 36 & Single-dish \\
104 & MAT80 & \citet{MAT80} & HMSFR & 1 & 0 & 25 & Interf. \\
105 & TUR72 & \citet{TUR72} & Discovery & 1 & 0 & 36 & Single-dish \\
106 & BAR71 & \citet{BAR71} & Discovery & 1 & 17 & 25 & Single-dish
\enddata
\end{deluxetable*}




\bibliography{maserdb}

\begin{thebibliography}{}
\expandafter\ifx\csname natexlab\endcsname\relax\def\natexlab#1{#1}\fi
\providecommand{\url}[1]{\href{#1}{#1}}

\bibitem[{{Araya} {et~al.}(2008){Araya}, {Hofner}, {Kurtz}, {Olmi}, \&
  {Linz}}]{ARA08}
{Araya}, E., {Hofner}, P., {Kurtz}, S., {Olmi}, L., \& {Linz}, H. 2008, \apj,
  675, 420

\bibitem[{{Araya} {et~al.}(2009){Araya}, {Kurtz}, {Hofner}, \& {Linz}}]{ARA09}
{Araya}, E.~D., {Kurtz}, S., {Hofner}, P., \& {Linz}, H. 2009, \apj, 698, 1321

\bibitem[{{Avedisova}(2002)}]{Avedisova2002}
{Avedisova}, V.~S. 2002, Astronomy Reports, 46, 193

\bibitem[{{Bachiller}(1996)}]{Bachiller1996}
{Bachiller}, R. 1996, \araa, 34, 111

\bibitem[{{Bachiller} {et~al.}(1990){Bachiller}, {Gomez-Gonzalez}, {Barcia}, \&
  {Menten}}]{BAC90}
{Bachiller}, R., {Gomez-Gonzalez}, J., {Barcia}, A., \& {Menten}, K.~M. 1990,
  \aap, 240, 116

\bibitem[{{Bae} {et~al.}(2011){Bae}, {Kim}, {Youn}, {Kim}, {Byun}, {Kang}, \&
  {Oh}}]{BAE11}
{Bae}, J.-H., {Kim}, K.-T., {Youn}, S.-Y., {et~al.} 2011, \apjs, 196, 21

\bibitem[{{Barrett} {et~al.}(1971{\natexlab{a}}){Barrett}, {Schwartz}, \&
  {Waters}}]{Barrett71}
{Barrett}, A.~H., {Schwartz}, P.~R., \& {Waters}, J.~W. 1971{\natexlab{a}},
  \apjl, 168, L101

\bibitem[{{Barrett} {et~al.}(1971{\natexlab{b}}){Barrett}, {Schwartz}, \&
  {Waters}}]{BAR71}
---. 1971{\natexlab{b}}, \apjl, 168, L101

\bibitem[{{Batrla} {et~al.}(1987){Batrla}, {Matthews}, {Menten}, \&
  {Walmsley}}]{Batrla1987}
{Batrla}, W., {Matthews}, H.~E., {Menten}, K.~M., \& {Walmsley}, C.~M. 1987,
  \nat, 326, 49

\bibitem[{{Batrla} \& {Menten}(1988)}]{BAT88}
{Batrla}, W., \& {Menten}, K.~M. 1988, \apj, 329, L117

\bibitem[{{Bayandina} {et~al.}(2012){Bayandina}, {Val'tts}, \&
  {Larionov}}]{Bayandina2012}
{Bayandina}, O.~S., {Val'tts}, I.~E., \& {Larionov}, G.~M. 2012, Astronomy
  Reports, 56, 553

\bibitem[{{Becker} {et~al.}(1991){Becker}, {White}, \& {Edwards}}]{Becker1991}
{Becker}, R.~H., {White}, R.~L., \& {Edwards}, A.~L. 1991, \apjs, 75, 1

\bibitem[{{Berulis} {et~al.}(1992){Berulis}, {Kalenski}, {Sobolev}, \&
  {Streinitski}}]{BER92}
{Berulis}, I.~I., {Kalenski}, S.~V., {Sobolev}, A.~M., \& {Streinitski}, V.~S.
  1992, Astronomical and Astrophysical Transactions, 1, 231

\bibitem[{{Beuther} {et~al.}(2005){Beuther}, {Thorwirth}, {Zhang}, {Hunter},
  {Megeath}, {Walsh}, \& {Menten}}]{BEU05}
{Beuther}, H., {Thorwirth}, S., {Zhang}, Q., {et~al.} 2005, \apj, 627, 834

\bibitem[{{Boch} \& {Fernique}(2014)}]{Boch2014}
{Boch}, T., \& {Fernique}, P. 2014, in Astronomical Society of the Pacific
  Conference Series, Vol. 485, Astronomical Data Analysis Software and Systems
  XXIII, ed. N.~{Manset} \& P.~{Forshay}, 277

\bibitem[{{Breen} {et~al.}(2019){Breen}, {Contreras}, {Dawson}, {Ellingsen},
  {Voronkov}, \& {McCarthy}}]{BRE19}
{Breen}, S.~L., {Contreras}, Y., {Dawson}, J.~R., {et~al.} 2019, \mnras, 484,
  5072

\bibitem[{{Breen} {et~al.}(2011){Breen}, {Ellingsen}, {Caswell}, {Green},
  {Fuller}, {Voronkov}, {Quinn}, \& {Avison}}]{Breen11}
{Breen}, S.~L., {Ellingsen}, S.~P., {Caswell}, J.~L., {et~al.} 2011, \apj, 733,
  80

\bibitem[{{Breen} {et~al.}(2015){Breen}, {Fuller}, {Caswell}, {Green},
  {Avison}, {Ellingsen}, {Gray}, {Pestalozzi}, {Quinn}, {Richards}, {Thompson},
  \& {Voronkov}}]{Breen15}
{Breen}, S.~L., {Fuller}, G.~A., {Caswell}, J.~L., {et~al.} 2015, \mnras, 450,
  4109

\bibitem[{{Bronfman} {et~al.}(2000){Bronfman}, {Casassus}, {May}, \&
  {Nyman}}]{Bronfman2000}
{Bronfman}, L., {Casassus}, S., {May}, J., \& {Nyman}, L.-{\AA}. 2000, \aap,
  358, 521

\bibitem[{{Chambers} {et~al.}(2011){Chambers}, {Yusef-Zadeh}, \&
  {Roberts}}]{CHA11}
{Chambers}, E.~T., {Yusef-Zadeh}, F., \& {Roberts}, D. 2011, \apj, 733, 42

\bibitem[{{Chen} {et~al.}(2015){Chen}, {Ellingsen}, {Baan}, {Qiao}, {Li}, {An},
  \& {Breen}}]{CHE15}
{Chen}, X., {Ellingsen}, S.~P., {Baan}, W.~A., {et~al.} 2015, \apj, 800, L2

\bibitem[{{Chen} {et~al.}(2019){Chen}, {Ellingsen}, {Ren}, {Sobolev},
  {Parfenov}, \& {Shen}}]{CHE19}
{Chen}, X., {Ellingsen}, S.~P., {Ren}, Z.-Y., {et~al.} 2019, \apj, 877, 90

\bibitem[{{Chen} {et~al.}(2018){Chen}, {Ellingsen}, {Shen}, {McCarthy},
  {Zhong}, \& {Deng}}]{CHE18}
{Chen}, X., {Ellingsen}, S.~P., {Shen}, Z.-Q., {et~al.} 2018, \apjl, 856, L35

\bibitem[{{Chen} {et~al.}(2011{\natexlab{a}}){Chen}, {Ellingsen}, {Shen},
  {Titmarsh}, \& {Gan}}]{CHE11}
{Chen}, X., {Ellingsen}, S.~P., {Shen}, Z.-Q., {Titmarsh}, A., \& {Gan}, C.-G.
  2011{\natexlab{a}}, \apjs, 196, 9

\bibitem[{{Chen} {et~al.}(2011{\natexlab{b}}){Chen}, {Ellingsen}, {Shen},
  {Titmarsh}, \& {Gan}}]{Chen2011}
---. 2011{\natexlab{b}}, \apjs, 196, 9

\bibitem[{{Chen} {et~al.}(2016){Chen}, {Ellingsen}, {Zhang}, {Wang}, {Shen},
  {Wu}, \& {Wu}}]{CHE16}
{Chen}, X., {Ellingsen}, S.~P., {Zhang}, J.~S., {et~al.} 2016, \mnras, 459, 357

\bibitem[{{Chen} {et~al.}(2012{\natexlab{a}}){Chen}, {Ellingsen}, {He}, {Xu},
  {Gan}, {Shen}, {An}, {Sun}, \& {Ju}}]{CHE12}
{Chen}, X., {Ellingsen}, S.~P., {He}, J.-H., {et~al.} 2012{\natexlab{a}},
  \apjs, 200, 5

\bibitem[{{Chen} {et~al.}(2012{\natexlab{b}}){Chen}, {Ellingsen}, {He}, {Xu},
  {Gan}, {Shen}, {An}, {Sun}, \& {Ju}}]{Chen2012}
---. 2012{\natexlab{b}}, \apjs, 200, 5

\bibitem[{{Chilingarian} {et~al.}(2004){Chilingarian}, {Bartunov}, {Richter},
  \& {Sigaev}}]{Chilingarian2004}
{Chilingarian}, I., {Bartunov}, O., {Richter}, J., \& {Sigaev}, T. 2004, in
  Astronomical Society of the Pacific Conference Series, Vol. 314, Astronomical
  Data Analysis Software and Systems (ADASS) XIII, ed. F.~{Ochsenbein}, M.~G.
  {Allen}, \& D.~{Egret}, 225

\bibitem[{{Choi} {et~al.}(2012){Choi}, {Kang}, {Byun}, \& {Lee}}]{CHO12}
{Choi}, M., {Kang}, M., {Byun}, D.-Y., \& {Lee}, J.-E. 2012, \apj, 759, 136

\bibitem[{{Condon} {et~al.}(1998){Condon}, {Cotton}, {Greisen}, {Yin},
  {Perley}, {Taylor}, \& {Broderick}}]{Condon1998}
{Condon}, J.~J., {Cotton}, W.~D., {Greisen}, E.~W., {et~al.} 1998, \aj, 115,
  1693

\bibitem[{{Cotton} \& {Yusef-Zadeh}(2016)}]{COT16}
{Cotton}, W.~D., \& {Yusef-Zadeh}, F. 2016, \apjs, 227, 10

\bibitem[{{Cragg} {et~al.}(2005){Cragg}, {Sobolev}, \& {Godfrey}}]{Cragg2005}
{Cragg}, D.~M., {Sobolev}, A.~M., \& {Godfrey}, P.~D. 2005, \mnras, 360, 533

\bibitem[{{Cyganowski} {et~al.}(2009){Cyganowski}, {Brogan}, {Hunter}, \&
  {Churchwell}}]{CYG09}
{Cyganowski}, C.~J., {Brogan}, C.~L., {Hunter}, T.~R., \& {Churchwell}, E.
  2009, \apj, 702, 1615

\bibitem[{{Cyganowski} {et~al.}(2008){Cyganowski}, {Whitney}, {Holden},
  {Braden}, {Brogan}, {Churchwell}, {Indebetouw}, {Watson}, {Babler},
  {Benjamin}, {Gomez}, {Meade}, {Povich}, {Robitaille}, \&
  {Watson}}]{Cyganowski2008}
{Cyganowski}, C.~J., {Whitney}, B.~A., {Holden}, E., {et~al.} 2008, \aj, 136,
  2391

\bibitem[{{Dame} {et~al.}(2001){Dame}, {Hartmann}, \& {Thaddeus}}]{Dame2001}
{Dame}, T.~M., {Hartmann}, D., \& {Thaddeus}, P. 2001, \apj, 547, 792

\bibitem[{{Darling} {et~al.}(2003){Darling}, {Goldsmith}, {Li}, \&
  {Giovanelli}}]{Darling2003}
{Darling}, J., {Goldsmith}, P., {Li}, D., \& {Giovanelli}, R. 2003, \aj, 125,
  1177

\bibitem[{{De Buizer} \& {Vacca}(2010)}]{DeBuizer2010}
{De Buizer}, J.~M., \& {Vacca}, W.~D. 2010, \aj, 140, 196

\bibitem[{{Deguchi} {et~al.}(2012){Deguchi}, {Tafoya}, \& {Nagisa}}]{DEG12B}
{Deguchi}, S., {Tafoya}, D., \& {Nagisa}, S. 2012, \pasj, 64, 28

\bibitem[{{Dobbs} \& {Burkert}(2012)}]{Dobbs2012}
{Dobbs}, C.~L., \& {Burkert}, A. 2012, \mnras, 421, 2940

\bibitem[{{Ellingsen}(2005)}]{ELL05}
{Ellingsen}, S.~P. 2005, \mnras, 359, 1498

\bibitem[{{Ellingsen} {et~al.}(2017){Ellingsen}, {Chen}, {Breen}, \&
  {Qiao}}]{ELL17}
{Ellingsen}, S.~P., {Chen}, X., {Breen}, S.~L., \& {Qiao}, H.~H. 2017, \mnras,
  472, 604

\bibitem[{{Ellingsen} {et~al.}(2014){Ellingsen}, {Chen}, {Qiao}, {Baan}, {An},
  {Li}, \& {Breen}}]{ELL14}
{Ellingsen}, S.~P., {Chen}, X., {Qiao}, H.-H., {et~al.} 2014, \apj, 790, L28

\bibitem[{{Ellingsen} {et~al.}(1994){Ellingsen}, {Norris}, {Whiteoak}, {Vaile},
  {McCullch}, \& {Price}}]{Ellingsen1994}
{Ellingsen}, S.~P., {Norris}, R.~P., {Whiteoak}, J.~B., {et~al.} 1994, \mnras,
  267, 510

\bibitem[{{Ellingsen} {et~al.}(2018){Ellingsen}, {Voronkov}, {Breen},
  {Caswell}, \& {Sobolev}}]{ELL18}
{Ellingsen}, S.~P., {Voronkov}, M.~A., {Breen}, S.~L., {Caswell}, J.~L., \&
  {Sobolev}, A.~M. 2018, \mnras, 480, 4851

\bibitem[{Ester {et~al.}(1996)Ester, Kriegel, Sander, \& Xu}]{Ester1996}
Ester, M., Kriegel, H.-P., Sander, J., \& Xu, X. 1996, in Proceedings of the
  Second International Conference on Knowledge Discovery and Data Mining,
  KDD'96 (AAAI Press), 226--231.
\newblock \url{http://dl.acm.org/citation.cfm?id=3001460.3001507}

\bibitem[{{Fish} {et~al.}(2011){Fish}, {Muehlbrad}, {Pratap}, {Sjouwerman},
  {Strelnitski}, {Pihlstr{\"o}m}, \& {Bourke}}]{FIS11}
{Fish}, V.~L., {Muehlbrad}, T.~C., {Pratap}, P., {et~al.} 2011, \apj, 729, 14

\bibitem[{{Fontani} {et~al.}(2010){Fontani}, {Cesaroni}, \& {Furuya}}]{FON10}
{Fontani}, F., {Cesaroni}, R., \& {Furuya}, R.~S. 2010, \aap, 517, A56

\bibitem[{{Forster} {et~al.}(1990){Forster}, {Caswell}, {Okumura}, {Hasegawa},
  \& {Ishiguro}}]{FOR90}
{Forster}, J.~R., {Caswell}, J.~L., {Okumura}, S.~K., {Hasegawa}, T., \&
  {Ishiguro}, M. 1990, \aap, 231, 473

\bibitem[{{Froebrich} {et~al.}(2011){Froebrich}, {Davis}, {Ioannidis},
  {Gledhill}, {Takami}, {Chrysostomou}, {Drew}, {Eisl{\"o}ffel}, {Gosling},
  {Gredel}, {Hatchell}, {Hodapp}, {Kumar}, {Lucas}, {Matthews}, {Rawlings},
  {Smith}, {Stecklum}, {Varricatt}, {Lee}, {Teixeira}, {Aspin}, {Khanzadyan},
  {Karr}, {Kim}, {Koo}, {Lee}, {Lee}, {Magakian}, {Movsessian}, {Nikogossian},
  {Pyo}, \& {Stanke}}]{Froebrich2011}
{Froebrich}, D., {Davis}, C.~J., {Ioannidis}, G., {et~al.} 2011, \mnras, 413,
  480

\bibitem[{{Gan} {et~al.}(2013){Gan}, {Chen}, {Shen}, {Xu}, \& {Ju}}]{GAN13}
{Gan}, C.-G., {Chen}, X., {Shen}, Z.-Q., {Xu}, Y., \& {Ju}, B.-G. 2013, \apj,
  763, 2

\bibitem[{{Ginsburg} {et~al.}(2013){Ginsburg}, {Glenn}, {Rosolowsky},
  {Ellsworth-Bowers}, {Battersby}, {Dunham}, {Merello}, {Shirley}, {Bally},
  {Evans}, {Stringfellow}, \& {Aguirre}}]{Ginsburg2013}
{Ginsburg}, A., {Glenn}, J., {Rosolowsky}, E., {et~al.} 2013, \apjs, 208, 14

\bibitem[{{G{\'o}mez} {et~al.}(2014){G{\'o}mez}, {Uscanga}, {Su{\'a}rez},
  {Rizzo}, \& {de Gregorio-Monsalvo}}]{GOM14}
{G{\'o}mez}, J.~F., {Uscanga}, L., {Su{\'a}rez}, O., {Rizzo}, J.~R., \& {de
  Gregorio-Monsalvo}, I. 2014, \rmxaa, 50, 137

\bibitem[{{G{\'o}mez} {et~al.}(2010){G{\'o}mez}, {Luis},
  {Hern{\'a}ndez-Curiel}, {Kurtz}, {Hofner}, \& {Araya}}]{GOM10}
{G{\'o}mez}, L., {Luis}, L., {Hern{\'a}ndez-Curiel}, I., {et~al.} 2010, \apjs,
  191, 207

\bibitem[{{G{\'o}mez-Ruiz} {et~al.}(2016){G{\'o}mez-Ruiz}, {Kurtz}, {Araya},
  {Hofner}, \& {Loinard}}]{GOM16}
{G{\'o}mez-Ruiz}, A.~I., {Kurtz}, S.~E., {Araya}, E.~D., {Hofner}, P., \&
  {Loinard}, L. 2016, \apjs, 222, 18

\bibitem[{{Gorski} {et~al.}(2017){Gorski}, {Ott}, {Rand}, {Meier}, {Momjian},
  \& {Schinnerer}}]{GOR17}
{Gorski}, M., {Ott}, J., {Rand}, R., {et~al.} 2017, \apj, 842, 124

\bibitem[{{Green} {et~al.}(2017){Green}, {Breen}, {Fuller},
  {McClure-Griffiths}, {Ellingsen}, {Voronkov}, {Avison}, {Brooks}, {Burton},
  {Chrysostomou}, {Cox}, {Diamond}, {Gray}, {Hoare}, {Masheder}, {Pestalozzi},
  {Phillips}, {Quinn}, {Richards}, {Thompson}, {Walsh}, {Ward-Thompson},
  {Wong-McSweeney}, \& {Yates}}]{Green17}
{Green}, J.~A., {Breen}, S.~L., {Fuller}, G.~A., {et~al.} 2017, \mnras, 469,
  1383

\bibitem[{{Haschick} \& {Baan}(1989)}]{HAS89}
{Haschick}, A.~D., \& {Baan}, W.~A. 1989, \apj, 339, 949

\bibitem[{{Haschick} {et~al.}(1990){Haschick}, {Menten}, \& {Baan}}]{HAS90}
{Haschick}, A.~D., {Menten}, K.~M., \& {Baan}, W.~A. 1990, \apj, 354, 556

\bibitem[{{Helou} \& {Walker}(1988)}]{Helou1988}
{Helou}, G., \& {Walker}, D.~W., eds. 1988, {Infrared astronomical satellite
  (IRAS) catalogs and atlases. Volume 7: The small scale structure catalog},
  Vol.~7, 1--265

\bibitem[{{Hunter} {et~al.}(2014){Hunter}, {Brogan}, {Cyganowski}, \&
  {Young}}]{HUN14}
{Hunter}, T.~R., {Brogan}, C.~L., {Cyganowski}, C.~J., \& {Young}, K.~H. 2014,
  \apj, 788, 187

\bibitem[{{Ishihara} {et~al.}(2010){Ishihara}, {Onaka}, {Kataza}, {Salama},
  {Alfageme}, {Cassatella}, {Cox}, {Garc{\'{\i}}a-Lario}, {Stephenson},
  {Cohen}, {Fujishiro}, {Fujiwara}, {Hasegawa}, {Ita}, {Kim}, {Matsuhara},
  {Murakami}, {M{\"u}ller}, {Nakagawa}, {Ohyama}, {Oyabu}, {Pyo}, {Sakon},
  {Shibai}, {Takita}, {Tanab{\'e}}, {Uemizu}, {Ueno}, {Usui}, {Wada},
  {Watarai}, {Yamamura}, \& {Yamauchi}}]{Ishihara2010}
{Ishihara}, D., {Onaka}, T., {Kataza}, H., {et~al.} 2010, \aap, 514, A1

\bibitem[{{Johnston} {et~al.}(1992){Johnston}, {Gaume}, {Stolovy}, {Wilson},
  {Walmsley}, \& {Menten}}]{JOH92}
{Johnston}, K.~J., {Gaume}, R., {Stolovy}, S., {et~al.} 1992, \apj, 385, 232

\bibitem[{{Jordan} {et~al.}(2017{\natexlab{a}}){Jordan}, {Walsh}, {Breen},
  {Ellingsen}, {Voronkov}, \& {Hyland}}]{JOR17}
{Jordan}, C.~H., {Walsh}, A.~J., {Breen}, S.~L., {et~al.} 2017{\natexlab{a}},
  \mnras, 471, 3915

\bibitem[{{Jordan} {et~al.}(2017{\natexlab{b}}){Jordan}, {Walsh}, {Breen},
  {Ellingsen}, {Voronkov}, \& {Hyland}}]{Jordan2017}
---. 2017{\natexlab{b}}, \mnras, 471, 3915

\bibitem[{{Jordan} {et~al.}(2015{\natexlab{a}}){Jordan}, {Walsh}, {Lowe},
  {Voronkov}, {Ellingsen}, {Breen}, {Purcell}, {Barnes}, {Burton},
  {Cunningham}, {Hill}, {Jackson}, {Longmore}, {Peretto}, \&
  {Urquhart}}]{Jordan2015}
{Jordan}, C.~H., {Walsh}, A.~J., {Lowe}, V., {et~al.} 2015{\natexlab{a}},
  \mnras, 448, 2344

\bibitem[{{Jordan} {et~al.}(2015{\natexlab{b}}){Jordan}, {Walsh}, {Lowe},
  {Voronkov}, {Ellingsen}, {Breen}, {Purcell}, {Barnes}, {Burton},
  {Cunningham}, {Hill}, {Jackson}, {Longmore}, {Peretto}, \&
  {Urquhart}}]{JOR15}
---. 2015{\natexlab{b}}, \mnras, 448, 2344

\bibitem[{{Kalenskii} {et~al.}(1994{\natexlab{a}}){Kalenskii}, {Berulis},
  {Val'tts}, {Dzura}, {Slysh}, \& {Vasil'kov}}]{KAL94}
{Kalenskii}, S.~V., {Berulis}, I.~I., {Val'tts}, I.~E., {et~al.}
  1994{\natexlab{a}}, \azh, 71, 51

\bibitem[{{Kalenskii} {et~al.}(2010{\natexlab{a}}){Kalenskii}, {Johansson},
  {Bergman}, {Kurtz}, {Hofner}, {Walmsley}, \& {Slysh}}]{KAL10B}
{Kalenskii}, S.~V., {Johansson}, L.~E.~B., {Bergman}, P., {et~al.}
  2010{\natexlab{a}}, \mnras, 405, 613

\bibitem[{{Kalenskii} {et~al.}(2013){Kalenskii}, {Kurtz}, \& {Bergman}}]{KAL13}
{Kalenskii}, S.~V., {Kurtz}, S., \& {Bergman}, P. 2013, Astronomy Reports, 57,
  120

\bibitem[{{Kalenskii} {et~al.}(2010{\natexlab{b}}){Kalenskii}, {Kurtz},
  {Slysh}, {Hofner}, {Walmsley}, {Johansson}, \& {Bergman}}]{KAL10}
{Kalenskii}, S.~V., {Kurtz}, S., {Slysh}, V.~I., {et~al.} 2010{\natexlab{b}},
  Astronomy Reports, 54, 932

\bibitem[{{Kalenskii} {et~al.}(1994{\natexlab{b}}){Kalenskii}, {Liljestroem},
  {Val'tts}, {Vasil'kov}, {Slysh}, \& {Urpo}}]{KAL94B}
{Kalenskii}, S.~V., {Liljestroem}, T., {Val'tts}, I.~E., {et~al.}
  1994{\natexlab{b}}, \aaps, 103, 129

\bibitem[{{Kalenskii} {et~al.}(1996){Kalenskii}, {Slysh}, {Val'tts}, \&
  {Dzura}}]{KAL96}
{Kalenskii}, S.~V., {Slysh}, V.~I., {Val'tts}, I.~E., \& {Dzura}, A.~M. 1996,
  in IAU Symposium, Vol. 170, CO: Twenty-Five Years of Millimeter-Wave
  Spectroscopy, 49

\bibitem[{{Kalenskij} {et~al.}(1992){Kalenskij}, {Bachiller}, {Berulis},
  {Val'tts}, {Gomez-Gonzalez}, {Martin-Pintado}, {Rodriguez-Franco}, \&
  {Slysh}}]{KAL92}
{Kalenskij}, S.~V., {Bachiller}, R., {Berulis}, I.~I., {et~al.} 1992, \azh, 69,
  1002

\bibitem[{{Kalenskij} {et~al.}(1993){Kalenskij}, {Berulis}, {Val'tts}, {Slysh},
  {Bachiller}, {Gomez-Gonzalez}, {Martin-Pintado}, \&
  {Rodriguez-Franco}}]{KAL93}
{Kalenskij}, S.~V., {Berulis}, I.~I., {Val'tts}, I.~E., {et~al.} 1993, {A study
  of methanol masers at 36 and 44 GHz and 48 GHz thermal emission around
  them.}, ed. A.~W. {Clegg} \& G.~E. {Nedoluha}, Vol. 412, 191--194

\bibitem[{{Kalenski{\u{i}}} {et~al.}(2006){Kalenski{\u{i}}}, {Promyslov},
  {Slysh}, {Bergman}, \& {Winnberg}}]{KAL06}
{Kalenski{\u{i}}}, S.~V., {Promyslov}, V.~G., {Slysh}, V.~I., {Bergman}, P., \&
  {Winnberg}, A. 2006, Astronomy Reports, 50, 289

\bibitem[{{Kalenski{\u{i}}} {et~al.}(2001){Kalenski{\u{i}}}, {Slysh},
  {Val'tts}, {Winnberg}, \& {Johansson}}]{KAL01}
{Kalenski{\u{i}}}, S.~V., {Slysh}, V.~I., {Val'tts}, I.~E., {Winnberg}, A., \&
  {Johansson}, L.~E. 2001, Astronomy Reports, 45, 26

\bibitem[{{Kang} {et~al.}(2015){Kang}, {Kim}, {Byun}, {Lee}, \& {Park}}]{KAN15}
{Kang}, H., {Kim}, K.-T., {Byun}, D.-Y., {Lee}, S., \& {Park}, Y.-S. 2015,
  \apjs, 221, 6

\bibitem[{{Kang} {et~al.}(2016){Kang}, {Byun}, {Kim}, {Kim}, {Lyo}, \&
  {Vlemmings}}]{KAN16}
{Kang}, J.-h., {Byun}, D.-Y., {Kim}, K.-T., {et~al.} 2016, \apjs, 227, 17

\bibitem[{{Kang} {et~al.}(2013){Kang}, {Lee}, {Choi}, {Choi}, {Kim}, {Di
  Francesco}, \& {Park}}]{KAN13}
{Kang}, M., {Lee}, J.-E., {Choi}, M., {et~al.} 2013, \apjs, 209, 25

\bibitem[{{Kim} {et~al.}(2019){Kim}, {Kim}, \& {Kim}}]{KIM19}
{Kim}, W.-J., {Kim}, K.-T., \& {Kim}, K.-T. 2019, The Astrophysical Journal
  Supplement Series, 244, 2

\bibitem[{{Kogan} \& {Slysh}(1998)}]{KOG98}
{Kogan}, L., \& {Slysh}, V. 1998, \apj, 497, 800

\bibitem[{{Kurtz} {et~al.}(2004{\natexlab{a}}){Kurtz}, {Hofner}, \&
  {{\'A}lvarez}}]{Kurtz2004}
{Kurtz}, S., {Hofner}, P., \& {{\'A}lvarez}, C.~V. 2004{\natexlab{a}}, \apjs,
  155, 149

\bibitem[{{Kurtz} {et~al.}(2004{\natexlab{b}}){Kurtz}, {Hofner}, \&
  {{\'A}lvarez}}]{KUR04}
---. 2004{\natexlab{b}}, \apjs, 155, 149

\bibitem[{{Ladeyschikov} {et~al.}(2016){Ladeyschikov}, {Kirsanova}, {Tsivilev},
  \& {Sobolev}}]{LAD16}
{Ladeyschikov}, D.~A., {Kirsanova}, M.~S., {Tsivilev}, A.~P., \& {Sobolev},
  A.~M. 2016, Astrophysical Bulletin, 71, 208

\bibitem[{{Li} {et~al.}(2017){Li}, {Xu}, {Chen}, {Lu}, {Sun}, {Du}, \&
  {Shen}}]{LI17}
{Li}, Y.-J., {Xu}, Y., {Chen}, X., {et~al.} 2017, Research in Astronomy and
  Astrophysics, 17, 125

\bibitem[{{Liechti} \& {Walmsley}(1997)}]{LIE97}
{Liechti}, S., \& {Walmsley}, C.~M. 1997, \aap, 321, 625

\bibitem[{{Liechti} \& {Wilson}(1996)}]{LIE96}
{Liechti}, S., \& {Wilson}, T.~L. 1996, \aap, 314, 615

\bibitem[{{Lim} {et~al.}(2012){Lim}, {Lyo}, {Kim}, \& {Byun}}]{LIM12}
{Lim}, W., {Lyo}, A.~R., {Kim}, K.-T., \& {Byun}, D.-Y. 2012, \aj, 144, 151

\bibitem[{{Litovchenko} {et~al.}(2011{\natexlab{a}}){Litovchenko}, {Alakoz},
  {Val'Tts}, \& {Larionov}}]{LIT11}
{Litovchenko}, I.~D., {Alakoz}, A.~V., {Val'Tts}, I.~E., \& {Larionov}, G.~M.
  2011{\natexlab{a}}, Astronomy Reports, 55, 978

\bibitem[{{Litovchenko} {et~al.}(2011{\natexlab{b}}){Litovchenko}, {Alakoz},
  {Val'Tts}, \& {Larionov}}]{LIT11B}
---. 2011{\natexlab{b}}, Astronomy Reports, 55, 1086

\bibitem[{{Lyo} {et~al.}(2014{\natexlab{a}}){Lyo}, {Kim}, {Byun}, \&
  {Lee}}]{LYO14}
{Lyo}, A.~R., {Kim}, J., {Byun}, D.-Y., \& {Lee}, H.-G. 2014{\natexlab{a}},
  \aj, 148, 80

\bibitem[{{Lyo} {et~al.}(2014{\natexlab{b}}){Lyo}, {Kim}, {Byun}, \&
  {Lee}}]{Lyo2014}
{Lyo}, A.-R., {Kim}, J., {Byun}, D.-Y., \& {Lee}, H.-G. 2014{\natexlab{b}},
  \aj, 148, 80

\bibitem[{{Matsakis} {et~al.}(1980){Matsakis}, {Wright}, {Townes}, {Welch},
  {Cheung}, \& {Askne}}]{MAT80}
{Matsakis}, D.~N., {Wright}, M.~C.~H., {Townes}, C.~H., {et~al.} 1980, \apj,
  236, 481

\bibitem[{{Matsumoto} {et~al.}(2014){Matsumoto}, {Hirota}, {Sugiyama}, {Kim},
  {Kim}, {Byun}, {Jung}, {Chibueze}, {Honma}, {Kameya}, {Kim}, {Lyo}, {Motogi},
  {Oh}, {Shino}, {Sunada}, {Bae}, {Chung}, {Chung}, {Cho}, {Han}, {Han},
  {Hwang}, {Je}, {Jike}, {Jung}, {Jung}, {Kang}, {Kang}, {Kang}, {Kan-ya},
  {Kawaguchi}, {Kim}, {Kim}, {Ryoung Kim}, {Kim}, {Kobayashi}, {Kono},
  {Kurayama}, {Lee}, {Lee}, {Lee}, {Lee}, {Lee}, {Lee}, {Minh}, {Miyazaki},
  {Oh}, {Oyama}, {Park}, {Roh}, {Sasao}, {Sawada-Satoh}, {Shibata}, {Sohn},
  {Song}, {Tamura}, {Wajima}, {Wi}, {Yeom}, \& {Yun}}]{MAT14}
{Matsumoto}, N., {Hirota}, T., {Sugiyama}, K., {et~al.} 2014, \apj, 789, L1

\bibitem[{{McCarthy} {et~al.}(2018{\natexlab{a}}){McCarthy}, {Ellingsen},
  {Breen}, {Henkel}, {Voronkov}, \& {Chen}}]{MCC18}
{McCarthy}, T.~P., {Ellingsen}, S.~P., {Breen}, S.~L., {et~al.}
  2018{\natexlab{a}}, \mnras, 480, 4578

\bibitem[{{McCarthy} {et~al.}(2018{\natexlab{b}}){McCarthy}, {Ellingsen},
  {Breen}, {Voronkov}, \& {Chen}}]{MCC18B}
{McCarthy}, T.~P., {Ellingsen}, S.~P., {Breen}, S.~L., {Voronkov}, M.~A., \&
  {Chen}, X. 2018{\natexlab{b}}, \apjl, 867, L4

\bibitem[{{McCarthy} {et~al.}(2017){McCarthy}, {Ellingsen}, {Chen}, {Breen},
  {Voronkov}, \& {Qiao}}]{MCC17}
{McCarthy}, T.~P., {Ellingsen}, S.~P., {Chen}, X., {et~al.} 2017, \apj, 846,
  156

\bibitem[{{McCarthy} {et~al.}(2018{\natexlab{c}}){McCarthy}, {Ellingsen},
  {Voronkov}, \& {Cim{\`o}}}]{MCC18C}
{McCarthy}, T.~P., {Ellingsen}, S.~P., {Voronkov}, M.~A., \& {Cim{\`o}}, G.
  2018{\natexlab{c}}, \mnras, 477, 507

\bibitem[{{McEwen} {et~al.}(2016){McEwen}, {Pihlstr{\"o}m}, \&
  {Sjouwerman}}]{MCE16B}
{McEwen}, B.~C., {Pihlstr{\"o}m}, Y.~M., \& {Sjouwerman}, L.~O. 2016, \apj,
  826, 189

\bibitem[{{Mehringer} \& {Menten}(1997)}]{MEH97}
{Mehringer}, D.~M., \& {Menten}, K.~M. 1997, \apj, 474, 346

\bibitem[{{Menten}(1991)}]{Menten1991}
{Menten}, K.~M. 1991, \apjl, 380, L75

\bibitem[{{Menten} {et~al.}(1986{\natexlab{a}}){Menten}, {Walmsley}, {Henkel},
  \& {Wilson}}]{MEN86}
{Menten}, K.~M., {Walmsley}, C.~M., {Henkel}, C., \& {Wilson}, T.~L.
  1986{\natexlab{a}}, \aap, 157, 318

\bibitem[{{Menten} {et~al.}(1986{\natexlab{b}}){Menten}, {Walmsley}, {Henkel},
  {Wilson}, {Snyder}, {Hollis}, \& {Lovas}}]{MEN86B}
{Menten}, K.~M., {Walmsley}, C.~M., {Henkel}, C., {et~al.} 1986{\natexlab{b}},
  \aap, 169, 271

\bibitem[{{Menten} {et~al.}(2009){Menten}, {Wilson}, {Leurini}, \&
  {Schilke}}]{Menten2009}
{Menten}, K.~M., {Wilson}, R.~W., {Leurini}, S., \& {Schilke}, P. 2009, \apj,
  692, 47

\bibitem[{{Momjian} \& {Sarma}(2017)}]{MOM17}
{Momjian}, E., \& {Sarma}, A.~P. 2017, \apj, 834, 168

\bibitem[{{Morimoto} {et~al.}(1985){Morimoto}, {Ohishi}, \& {Kanzawa}}]{MOR85}
{Morimoto}, M., {Ohishi}, M., \& {Kanzawa}, T. 1985, \apj, 288, L11

\bibitem[{{M{\"u}ller} {et~al.}(2004){M{\"u}ller}, {Menten}, \&
  {M{\"a}der}}]{Muller2004}
{M{\"u}ller}, H.~S.~P., {Menten}, K.~M., \& {M{\"a}der}, H. 2004, \aap, 428,
  1019

\bibitem[{{M{\"u}ller} {et~al.}(2005){M{\"u}ller}, {Schl{\"o}der}, {Stutzki},
  \& {Winnewisser}}]{Muller2005}
{M{\"u}ller}, H.~S.~P., {Schl{\"o}der}, F., {Stutzki}, J., \& {Winnewisser}, G.
  2005, Journal of Molecular Structure, 742, 215

\bibitem[{{Nakano} \& {Yoshida}(1986)}]{NAK86}
{Nakano}, M., \& {Yoshida}, S. 1986, \pasj, 38, 531

\bibitem[{{Nakashima} {et~al.}(2015){Nakashima}, {Sobolev}, {Salii}, {Zhang},
  {Yung}, \& {Deguchi}}]{NAK15}
{Nakashima}, J.-i., {Sobolev}, A.~M., {Salii}, S.~V., {et~al.} 2015, \pasj, 67,
  95

\bibitem[{{Pandian} \& {Goldsmith}(2007)}]{Pandian07}
{Pandian}, J.~D., \& {Goldsmith}, P.~F. 2007, \apj, 669, 435

\bibitem[{Pedregosa {et~al.}(2012)Pedregosa, Varoquaux, Gramfort, Michel,
  Thirion, Grisel, Blondel, Prettenhofer, Weiss, Dubourg, Vanderplas, Passos,
  Cournapeau, Brucher, Perrot, Duchesnay, \& Louppe}]{Pedregosa2012}
Pedregosa, F., Varoquaux, G., Gramfort, A., {et~al.} 2012, Journal of Machine
  Learning Research, 12

\bibitem[{{Pestalozzi} {et~al.}(2007){Pestalozzi}, {Chrysostomou}, {Collett},
  {Minier}, {Conway}, \& {Booth}}]{Pestalozzi07}
{Pestalozzi}, M.~R., {Chrysostomou}, A., {Collett}, J.~L., {et~al.} 2007, \aap,
  463, 1009

\bibitem[{{Pestalozzi} {et~al.}(2005){Pestalozzi}, {Minier}, \&
  {Booth}}]{Pestalozzi05}
{Pestalozzi}, M.~R., {Minier}, V., \& {Booth}, R.~S. 2005, \aap, 432, 737

\bibitem[{{Phillips} {et~al.}(1998){Phillips}, {Norris}, {Ellingsen}, \&
  {Rayner}}]{Phillips1998}
{Phillips}, C.~J., {Norris}, R.~P., {Ellingsen}, S.~P., \& {Rayner}, D.~P.
  1998, \mnras, 294, 265

\bibitem[{{Pihlstr{\"o}m} {et~al.}(2011){Pihlstr{\"o}m}, {Sjouwerman}, \&
  {Fish}}]{PIH11}
{Pihlstr{\"o}m}, Y.~M., {Sjouwerman}, L.~O., \& {Fish}, V.~L. 2011, \apj, 739,
  L21

\bibitem[{{Pihlstr{\"o}m} {et~al.}(2014{\natexlab{a}}){Pihlstr{\"o}m},
  {Sjouwerman}, {Frail}, {Claussen}, {Mesler}, \& {McEwen}}]{Pihlstrom2014}
{Pihlstr{\"o}m}, Y.~M., {Sjouwerman}, L.~O., {Frail}, D.~A., {et~al.}
  2014{\natexlab{a}}, \aj, 147, 73

\bibitem[{{Pihlstr{\"o}m} {et~al.}(2014{\natexlab{b}}){Pihlstr{\"o}m},
  {Sjouwerman}, {Frail}, {Claussen}, {Mesler}, \& {McEwen}}]{PIH14}
---. 2014{\natexlab{b}}, \aj, 147, 73

\bibitem[{{Pineau} {et~al.}(2011){Pineau}, {Boch}, \& {Derriere}}]{Pineau2011}
{Pineau}, F.-X., {Boch}, T., \& {Derriere}, S. 2011, in Astronomical Society of
  the Pacific Conference Series, Vol. 442, Astronomical Data Analysis Software
  and Systems XX, ed. I.~N. {Evans}, A.~{Accomazzi}, D.~J. {Mink}, \& A.~H.
  {Rots}, 85

\bibitem[{{Plambeck} \& {Menten}(1990)}]{PLA90}
{Plambeck}, R.~L., \& {Menten}, K.~M. 1990, \apj, 364, 555

\bibitem[{{Polushkin} \& {Val'Tts}(2010)}]{POL10}
{Polushkin}, S.~V., \& {Val'Tts}, I.~E. 2010, Astronomy Reports, 54, 496

\bibitem[{{Rodr{\'\i}guez-Garza} {et~al.}(2017){Rodr{\'\i}guez-Garza}, {Kurtz},
  {G{\'o}mez-Ruiz}, {Hofner}, {Araya}, \& {Kalenskii}}]{ROD17B}
{Rodr{\'\i}guez-Garza}, C.~B., {Kurtz}, S.~E., {G{\'o}mez-Ruiz}, A.~I.,
  {et~al.} 2017, \apjs, 233, 4

\bibitem[{{R{\"o}llig} {et~al.}(1999){R{\"o}llig}, {Kegel}, {Mauersberger}, \&
  {Doerr}}]{ROL99}
{R{\"o}llig}, M., {Kegel}, W.~H., {Mauersberger}, R., \& {Doerr}, C. 1999,
  \aap, 343, 939

\bibitem[{{Salii} {et~al.}(2002){Salii}, {Sobolev}, \& {Kalinina}}]{Salii2002}
{Salii}, S.~V., {Sobolev}, A.~M., \& {Kalinina}, N.~D. 2002, Astronomy Reports,
  46, 955

\bibitem[{{Sinclair} {et~al.}(1992){Sinclair}, {Carrad}, {Caswell}, {Norris},
  \& {Whiteoak}}]{Sinclair1992}
{Sinclair}, M.~W., {Carrad}, G.~J., {Caswell}, J.~L., {Norris}, R.~P., \&
  {Whiteoak}, J.~B. 1992, \mnras, 256, 33P

\bibitem[{{Sjouwerman} {et~al.}(2011){Sjouwerman}, {Pihlstr{\"o}m}, \&
  {Fish}}]{SJO11}
{Sjouwerman}, L.~O., {Pihlstr{\"o}m}, Y.~M., \& {Fish}, V.~L. 2011, in
  Astronomical Society of the Pacific Conference Series, Vol. 439, The Galactic
  Center: a Window to the Nuclear Environment of Disk Galaxies, ed. M.~R.
  {Morris}, Q.~D. {Wang}, \& F.~{Yuan}, 75

\bibitem[{{Slysh} \& {Kalenskii}(2009)}]{SLY09}
{Slysh}, V.~I., \& {Kalenskii}, S.~V. 2009, Astronomy Reports, 53, 519

\bibitem[{{Slysh} {et~al.}(1997){Slysh}, {Kalenskii}, {Val'tts}, \&
  {Golubev}}]{SLY97}
{Slysh}, V.~I., {Kalenskii}, S.~V., {Val'tts}, I.~E., \& {Golubev}, V.~V. 1997,
  \apj, 478, L37

\bibitem[{{Slysh} {et~al.}(1999){Slysh}, {Kalenskii}, {Val'TTS}, {Golubev}, \&
  {Mead}}]{SLY99}
{Slysh}, V.~I., {Kalenskii}, S.~V., {Val'TTS}, I.~E., {Golubev}, V.~V., \&
  {Mead}, K. 1999, \apjs, 123, 515

\bibitem[{{Slysh} {et~al.}(1994){Slysh}, {Kalenskii}, {Valtts}, \&
  {Otrupcek}}]{SLY94}
{Slysh}, V.~I., {Kalenskii}, S.~V., {Valtts}, I.~E., \& {Otrupcek}, R. 1994,
  \mnras, 268, 464

\bibitem[{{Slysh} {et~al.}(1993){Slysh}, {Kalenskij}, \& {Val'tts}}]{SLY93}
{Slysh}, V.~I., {Kalenskij}, S.~V., \& {Val'tts}, I.~E. 1993, \apj, 413, L133

\bibitem[{{Slysh} {et~al.}(2002){Slysh}, {Kalenski{\u{i}}}, \&
  {Val'tts}}]{SLY02}
{Slysh}, V.~I., {Kalenski{\u{i}}}, S.~V., \& {Val'tts}, I.~E. 2002, Astronomy
  Reports, 46, 49

\bibitem[{{Sobolev}(1993)}]{Sobolev1993}
{Sobolev}, A.~M. 1993, Astronomy Letters, 19, 293

\bibitem[{{Sobolev} {et~al.}(2005){Sobolev}, {Ostrovskii}, {Kirsanova},
  {Shelemei}, {Voronkov}, \& {Malyshev}}]{Sobolev2005}
{Sobolev}, A.~M., {Ostrovskii}, A.~B., {Kirsanova}, M.~S., {et~al.} 2005, in
  IAU Symposium, Vol. 227, Massive Star Birth: A Crossroads of Astrophysics,
  ed. R.~{Cesaroni}, M.~{Felli}, E.~{Churchwell}, \& M.~{Walmsley}, 174--179

\bibitem[{{Sobolev} {et~al.}(2007){Sobolev}, {Cragg}, {Ellingsen}, {Gaylard},
  {Goedhart}, {Henkel}, {Kirsanova}, {Ostrovskii}, {Pankratova}, {Shelemei},
  {van der Walt}, {Vasyunina}, \& {Voronkov}}]{Sobolev2007}
{Sobolev}, A.~M., {Cragg}, D.~M., {Ellingsen}, S.~P., {et~al.} 2007, in IAU
  Symposium, Vol. 242, Astrophysical Masers and their Environments, ed. J.~M.
  {Chapman} \& W.~A. {Baan}, 81--88

\bibitem[{{Szymczak} {et~al.}(2012){Szymczak}, {Wolak}, {Bartkiewicz}, \&
  {Borkowski}}]{Szymczak12}
{Szymczak}, M., {Wolak}, P., {Bartkiewicz}, A., \& {Borkowski}, K.~M. 2012,
  Astronomische Nachrichten, 333, 634

\bibitem[{{Taylor}(2005)}]{Taylor2005}
{Taylor}, M.~B. 2005, in Astronomical Society of the Pacific Conference Series,
  Vol. 347, Astronomical Data Analysis Software and Systems XIV, ed.
  P.~{Shopbell}, M.~{Britton}, \& R.~{Ebert} (Pasadena, California, USA:
  Astronomical Society of the Pacific), 29

\bibitem[{{Towner} {et~al.}(2017){Towner}, {Brogan}, {Hunter}, {Cyganowski},
  {McGuire}, {Indebetouw}, {Friesen}, \& {Chandler}}]{TOW17}
{Towner}, A.~P.~M., {Brogan}, C.~L., {Hunter}, T.~R., {et~al.} 2017, \apjs,
  230, 22

\bibitem[{{Turner} {et~al.}(1972{\natexlab{a}}){Turner}, {Gordon}, \&
  {Wrixon}}]{Turner72}
{Turner}, B.~E., {Gordon}, M.~A., \& {Wrixon}, G.~T. 1972{\natexlab{a}}, \apj,
  177, 609

\bibitem[{{Turner} {et~al.}(1972{\natexlab{b}}){Turner}, {Gordon}, \&
  {Wrixon}}]{TUR72}
---. 1972{\natexlab{b}}, \apj, 177, 609

\bibitem[{{Val'tts}(1998)}]{VAL98}
{Val'tts}, I.~E. 1998, Astronomy Letters, 24, 788

\bibitem[{{Val'tts} {et~al.}(1995){Val'tts}, {Dzyura}, {Kalenskii}, {Slysh},
  {Bus}, \& {Vinnberg}}]{VAL95}
{Val'tts}, I.~E., {Dzyura}, A.~M., {Kalenskii}, S.~V., {et~al.} 1995, \azh, 72,
  22

\bibitem[{{Val'tts} {et~al.}(2000){Val'tts}, {Ellingsen}, {Slysh}, {Kalenskii},
  {Otrupcek}, \& {Larionov}}]{VAL00}
{Val'tts}, I.~E., {Ellingsen}, S.~P., {Slysh}, V.~I., {et~al.} 2000, \mnras,
  317, 315

\bibitem[{{Val'tts} \& {Larionov}(2007)}]{Valtts2007}
{Val'tts}, I.~E., \& {Larionov}, G.~M. 2007, Astronomy Reports, 51, 519

\bibitem[{{Val'tts} {et~al.}(2010){Val'tts}, {Larionov}, \&
  {Bayandina}}]{Valtts2010}
{Val'tts}, I.~E., {Larionov}, G.~M., \& {Bayandina}, O.~S. 2010, arXiv
  e-prints, arXiv:1005.3715

\bibitem[{{Varricatt} {et~al.}(2010){Varricatt}, {Davis}, {Ramsay}, \&
  {Todd}}]{Varricatt2010}
{Varricatt}, W.~P., {Davis}, C.~J., {Ramsay}, S., \& {Todd}, S.~P. 2010,
  \mnras, 404, 661

\bibitem[{{Voronkov} {et~al.}(2005{\natexlab{a}}){Voronkov}, {Sobolev},
  {Ellingsen}, {Ostrovskii}, \& {Alakoz}}]{VOR04}
{Voronkov}, M., {Sobolev}, A., {Ellingsen}, S., {Ostrovskii}, A., \& {Alakoz},
  A. 2005{\natexlab{a}}, \apss, 295, 217

\bibitem[{{Voronkov} {et~al.}(2006{\natexlab{a}}){Voronkov}, {Brooks},
  {Sobolev}, {Ellingsen}, {Ostrovskii}, \& {Caswell}}]{Voronkov2006}
{Voronkov}, M.~A., {Brooks}, K.~J., {Sobolev}, A.~M., {et~al.}
  2006{\natexlab{a}}, \mnras, 373, 411

\bibitem[{{Voronkov} {et~al.}(2006{\natexlab{b}}){Voronkov}, {Brooks},
  {Sobolev}, {Ellingsen}, {Ostrovskii}, \& {Caswell}}]{VOR06}
---. 2006{\natexlab{b}}, \mnras, 373, 411

\bibitem[{{Voronkov} {et~al.}(2010{\natexlab{a}}){Voronkov}, {Caswell},
  {Britton}, {Green}, {Sobolev}, \& {Ellingsen}}]{Voronkov2010a}
{Voronkov}, M.~A., {Caswell}, J.~L., {Britton}, T.~R., {et~al.}
  2010{\natexlab{a}}, \mnras, 408, 133

\bibitem[{{Voronkov} {et~al.}(2010{\natexlab{b}}){Voronkov}, {Caswell},
  {Britton}, {Green}, {Sobolev}, \& {Ellingsen}}]{VOR10}
---. 2010{\natexlab{b}}, \mnras, 408, 133

\bibitem[{{Voronkov} {et~al.}(2014{\natexlab{a}}){Voronkov}, {Caswell},
  {Ellingsen}, {Green}, \& {Breen}}]{Voronkov2014}
{Voronkov}, M.~A., {Caswell}, J.~L., {Ellingsen}, S.~P., {Green}, J.~A., \&
  {Breen}, S.~L. 2014{\natexlab{a}}, \mnras, 439, 2584

\bibitem[{{Voronkov} {et~al.}(2014{\natexlab{b}}){Voronkov}, {Caswell},
  {Ellingsen}, {Green}, \& {Breen}}]{VOR14}
---. 2014{\natexlab{b}}, \mnras, 439, 2584

\bibitem[{{Voronkov} {et~al.}(2010{\natexlab{c}}){Voronkov}, {Caswell},
  {Ellingsen}, \& {Sobolev}}]{Voronkov2010b}
{Voronkov}, M.~A., {Caswell}, J.~L., {Ellingsen}, S.~P., \& {Sobolev}, A.~M.
  2010{\natexlab{c}}, \mnras, 405, 2471

\bibitem[{{Voronkov} {et~al.}(2010{\natexlab{d}}){Voronkov}, {Caswell},
  {Ellingsen}, \& {Sobolev}}]{VOR10B}
---. 2010{\natexlab{d}}, \mnras, 405, 2471

\bibitem[{{Voronkov} {et~al.}(2005{\natexlab{b}}){Voronkov}, {Sobolev},
  {Ellingsen}, \& {Ostrovskii}}]{VOR05}
{Voronkov}, M.~A., {Sobolev}, A.~M., {Ellingsen}, S.~P., \& {Ostrovskii}, A.~B.
  2005{\natexlab{b}}, \mnras, 362, 995

\bibitem[{{Voronkov} {et~al.}(2011){Voronkov}, {Walsh}, {Caswell}, {Ellingsen},
  {Breen}, {Longmore}, {Purcell}, \& {Urquhart}}]{VOR11}
{Voronkov}, M.~A., {Walsh}, A.~J., {Caswell}, J.~L., {et~al.} 2011, \mnras,
  413, 2339

\bibitem[{{Walsh} {et~al.}(2011){Walsh}, {Breen}, {Britton}, {Brooks},
  {Burton}, {Cunningham}, {Green}, {Harvey-Smith}, {Hindson}, {Hoare},
  {Indermuehle}, {Jones}, {Lo}, {Longmore}, {Lowe}, {Phillips}, {Purcell},
  {Thompson}, {Urquhart}, {Voronkov}, {White}, \& {Whiting}}]{Walsh2011}
{Walsh}, A.~J., {Breen}, S.~L., {Britton}, T., {et~al.} 2011, \mnras, 416, 1764

\bibitem[{{Wang} {et~al.}(2014){Wang}, {Zhang}, {Gao}, {Zhang}, {Li}, {Fang},
  \& {Shi}}]{WAN14}
{Wang}, J., {Zhang}, J., {Gao}, Y., {et~al.} 2014, Nature Communications, 5,
  5449

\bibitem[{{White} \& {Becker}(1992)}]{White1992}
{White}, R.~L., \& {Becker}, R.~H. 1992, \apjs, 79, 331

\bibitem[{{Wiesemeyer} {et~al.}(2004){Wiesemeyer}, {Thum}, \&
  {Walmsley}}]{WIE04}
{Wiesemeyer}, H., {Thum}, C., \& {Walmsley}, C.~M. 2004, \aap, 428, 479

\bibitem[{{Wright} \& {Otrupcek}(1990)}]{Wright1990}
{Wright}, A., \& {Otrupcek}, R. 1990, in PKS Catalog (1990)

\bibitem[{{Wright} {et~al.}(1994){Wright}, {Griffith}, {Burke}, \&
  {Ekers}}]{Wright1994}
{Wright}, A.~E., {Griffith}, M.~R., {Burke}, B.~F., \& {Ekers}, R.~D. 1994,
  \apjs, 91, 111

\bibitem[{{Wu} {et~al.}(2004){Wu}, {Wei}, {Zhao}, {Shi}, {Yu}, {Qin}, \&
  {Huang}}]{Wu2004}
{Wu}, Y., {Wei}, Y., {Zhao}, M., {et~al.} 2004, \aap, 426, 503

\bibitem[{{Yanagida} {et~al.}(2014){Yanagida}, {Sakai}, {Hirota}, {Sakai},
  {Foster}, {Sanhueza}, {Jackson}, {Furuya}, {Aikawa}, \& {Yamamoto}}]{YAN14}
{Yanagida}, T., {Sakai}, T., {Hirota}, T., {et~al.} 2014, \apjl, 794, L10

\bibitem[{{Yang} {et~al.}(2017{\natexlab{a}}){Yang}, {Chen}, {Shen}, {Li},
  {Wang}, {Jiang}, {Li}, {Dong}, {Wu}, {Qiao}, \& {Ren}}]{Yang17}
{Yang}, K., {Chen}, X., {Shen}, Z.-Q., {et~al.} 2017{\natexlab{a}}, \apj, 846,
  160

\bibitem[{{Yang} {et~al.}(2019){Yang}, {Chen}, {Shen}, {Li}, {Wang}, {Jiang},
  {Li}, {Dong}, {Wu}, \& {Qiao}}]{Yang19}
---. 2019, \apjs, 241, 18

\bibitem[{{Yang} {et~al.}(2017{\natexlab{b}}){Yang}, {Xu}, {Chen}, {Ellingsen},
  {Lu}, {Ju}, \& {Li}}]{YAN17}
{Yang}, W., {Xu}, Y., {Chen}, X., {et~al.} 2017{\natexlab{b}}, \apjs, 231, 20

\bibitem[{{Yang} {et~al.}(2017{\natexlab{c}}){Yang}, {Xu}, {Chen}, {Ellingsen},
  {Lu}, {Ju}, \& {Li}}]{Yang2017}
---. 2017{\natexlab{c}}, \apjs, 231, 20

\bibitem[{{Yusef-Zadeh} {et~al.}(2013{\natexlab{a}}){Yusef-Zadeh}, {Cotton},
  {Viti}, {Wardle}, \& {Royster}}]{YUS13}
{Yusef-Zadeh}, F., {Cotton}, W., {Viti}, S., {Wardle}, M., \& {Royster}, M.
  2013{\natexlab{a}}, \apjl, 764, L19

\bibitem[{{Yusef-Zadeh} {et~al.}(2013{\natexlab{b}}){Yusef-Zadeh}, {Cotton},
  {Viti}, {Wardle}, \& {Royster}}]{Yusef2013}
---. 2013{\natexlab{b}}, \apjl, 764, L19

\bibitem[{{Zubrin} {et~al.}(2007){Zubrin}, {Antyufeyev}, {Myshenko}, \&
  {Shulga}}]{ZUB07}
{Zubrin}, S.~Y., {Antyufeyev}, A.~V., {Myshenko}, V.~V., \& {Shulga}, V.~M.
  2007, in 14th Young Scientists Conference on Astronomy and Space Physics, ed.
  G.~{Ivashchenko} \& A.~{Golovin}, 95--98

\bibitem[{{Zubrin} \& {Shulga}(2008)}]{ZUB08}
{Zubrin}, S.~Y., \& {Shulga}, V.~M. 2008, in Young Scientists 15th Proceedings,
  ed. V.~Y. {Choliy} \& G.~{Ivashchenko}, 41--43

\end{thebibliography}



\end{document}